\documentclass[]{scrartcl}
\usepackage{thumbpdf,lmodern}

\usepackage[utf8]{inputenc}

\title{intRinsic: an R Package for Model-Based Estimation of the
	Intrinsic Dimension of a Dataset}
\author{Francesco Denti\\Università Cattolica
	del Sacro Cuore}
\date{}

\usepackage{booktabs}
\usepackage{longtable}
\usepackage{array}
\usepackage{multirow}
\usepackage{wrapfig}
\usepackage{float}
\usepackage{colortbl}
\usepackage{pdflscape}
\usepackage{tabu}
\usepackage{threeparttable}
\usepackage{threeparttablex}
\usepackage[normalem]{ulem}
\usepackage{makecell}
\usepackage{natbib}
\usepackage{xcolor}
\usepackage{tikz}
\usepackage{amsmath,amsthm,amsfonts} \usepackage{placeins}   \usepackage{bm}   \usepackage{comment}   \usepackage{bbm,dsfont}   \usepackage{verbatim}   \usepackage{tabularx}   \usepackage{booktabs}   \usepackage{graphicx}   \usepackage{pifont}                  \usepackage{amssymb}  
\usepackage{hyperref}

\begin{document}
	
\maketitle

\begin{abstract}
This article illustrates \texttt{intRinsic}, an \texttt{R} package
that implements novel state-of-the-art likelihood-based estimators of
the intrinsic dimension of a dataset, an essential quantity for most
dimensionality reduction techniques. In order to make these novel
estimators easily accessible, the package contains a small number of
high-level functions that rely on a broader set of efficient, low-level
routines. Generally speaking, \texttt{intRinsic} encompasses models that
fall into two categories: homogeneous and heterogeneous intrinsic
dimension estimators. The first category contains the
\emph{two nearest neighbors} estimator, a method derived from the
distributional properties of the ratios of the distances between each
data point and its first two closest neighbors. The functions dedicated
to this method carry out inference under both the frequentist and
Bayesian frameworks. In the second category, we find the
\emph{heterogeneous intrinsic dimension algorithm}, a Bayesian mixture
model for which an efficient Gibbs sampler is implemented. After
presenting the theoretical background, we demonstrate the performance of
the models on simulated datasets. This way, we can facilitate the
exposition by immediately assessing the validity of the results. Then,
we employ the package to study the intrinsic dimension of the
\emph{Alon} dataset, obtained from a famous microarray experiment.
Finally, we show how the estimation of homogeneous and heterogeneous
intrinsic dimensions allows us to gain valuable insights into the
topological structure of a dataset.\\

\end{abstract}

\section{Introduction}
\label{sec:intro}

Statisticians and data scientists are often called to manipulate,
analyze, and summarize datasets that present high-dimensional and
elaborate dependency structures. In numerous cases, these large datasets
contain variables characterized by a considerable amount of redundant
information. One can exploit these redundancies to represent a large
dataset on a much lower-dimensional scale. This summarization procedure,
called \emph{dimensionality reduction}, is a fundamental step in many
statistical analyses. For example, dimensionality reduction techniques
grant the feasibility of otherwise challenging tasks such as large data
manipulation and visualization by reducing computational time and memory
requirements.

More formally, dimensionality reduction is possible whenever the data
points take place on one or more manifolds characterized by a lower
dimension than what has been observed initially. In this context, the
word \emph{manifold} is used to indicate a constraint surface embedded
in high-dimensional space along which dissimilarities between data
points are best represented \citep{Tenn, Lee2008}. We call the dimension
of a latent, potentially nonlinear manifold the
\emph{intrinsic dimension} (ID). Several other definitions of ID exist
in the literature. For example, we can regard the ID as the minimal
number of parameters needed to represent all the information contained
in the data without significant information loss
\citep{Ansuini2019, Rozza2011, Bennett1969}.\\
Intuitively, the ID is an indicator of the complexity of the features of
a dataset. It is a necessary piece of information to have before
attempting to perform any dimensionality reduction, manifold learning,
or visualization tasks. Indeed, most dimensionality reduction methods
would be worthless without a reliable estimate of the true ID they need
to target: an underestimated ID value can cause needless information
loss. At the same time, the reverse can lead to an unnecessary waste of
time and computational resources \citep{Hino2017}. Beyond dimensionality
reduction, ID estimation methods have been successfully employed, for
instance, in studying physical systems \citep{Mendes-Santos2021} and
analyzing neural networks \citep{Ansuini2019}. For more examples of
applications, see also \citet{Carter2010} and the references in the
discussion in \citet{Bac2021}.

Over the past few decades, a vast number of methods for ID estimation
and dimensionality reduction have been developed. The algorithms can be
broadly classified into two main categories: projection and geometric
approaches. The former maps the original data to a lower-dimensional
space. The projection function can be linear, as in the case of
Principal Component Analysis (PCA) \citep{Hotelling1933} or nonlinear,
as in the case of Locally Linear Embedding \citep{Roweis2000}, Isomap
\citep{Tenn}, and the tSNE \citep{LaurensvanderMaatenTiCC2009}. For more
examples, see \citet{Jollife2016} and the references therein. In
consequence, there is a plethora of \texttt{R} packages that implement
these types of algorithms. To mention some examples, one can use the
packages \texttt{RDRToolbox} \citep{RDRtoolbox}, \texttt{lle}
\citep{Kayo2006}, \texttt{Rtsne} \citep{Rtsne}, and the classic
\texttt{princomp()} function from the default package \texttt{stats}
\citep{Rcore}.

Instead, geometric approaches rely on the topology of a dataset,
exploiting the properties of the distances between data points. Within
this family, we can find fractal methods \citep{Falconer2003}, graphical
methods \citep{Costa2004}, model-based likelihood approaches
\citep{Levina}, and methods based on nearest neighbors distances
\citep{Pettis1979}. Also in this case, numerous packages are available:
for example, for fractal methods alone there are \texttt{fractaldim}
\citep{fractaldimpack}, \texttt{nonlinearTseries} \citep{nonlinearTSpack},
and \texttt{tseriesChaos} \citep{tserieschaospack}, among others. For a
recent review of the methodologies used for ID estimation we refer to
\citet{Campadelli2015}.

Given the abundance of approaches in this area, several \texttt{R}
developers have also attempted to provide unifying collections of
dimensionality reduction and ID estimation techniques. For example,
remarkable ensembles of methodologies are implemented in the packages
\texttt{ider} \citep{Hino2017b}, \texttt{dimred} and \texttt{coRanking}
\citep{dimred}, \texttt{dyndimred} \citep{dyndimred}, \texttt{IDmining}
\citep{Golay2017}, and \texttt{intrinsicDimension}
\citep{intrinsicDimension}. Among the various options, the package
\texttt{Rdimtools} \citep{rdimtoolspack} stands out, implementing 150
different algorithms, 17 of which are exclusively dedicated to ID
estimation \citep{You2020}. Finally, it is worth mentioning that there
are also \texttt{Python} packages implementing different methods for
ID estimation: two prominent examples are \texttt{scikit-learn}
\citep{Bac2021} and \texttt{DADApy} \citep{dadapy}. See Section B of the
Appendix for more details.

In this paper, we introduce and discuss the \texttt{R} package
\texttt{intRinsic} (version 0.2.2). The package is openly available from
the Comprehensive \texttt{R} Archive Network (CRAN) at
\url{https://CRAN.R-project.org/package=intRinsic}, and can be installed
by running

\begin{verbatim}
R> install.packages("intRinsic")
\end{verbatim}

Future developments and updates will be uploaded both on CRAN and GitHub
at \url{https://github.com/Fradenti/intRinsic}.

The package implements the two nearest neighbors (TWO-NN), the
generalized ratio ID estimator (GRIDE), and the heterogeneous ID
algorithm (HIDALGO) models, three state-of-the-art ID estimators
recently introduced in \citet{Facco,Denti2021} and \citet{Allegra},
respectively. These methods are likelihood-based estimators that rely on
the theoretical properties of the distances among nearest neighbors. The
first two models estimate a global, unique ID of a dataset and are
implemented under both the frequentist and Bayesian paradigms. Moreover,
one can exploit these models to study how the ID depends on the scale of
the neighborhood considered for its estimation. On the contrary, HIDALGO
is a Bayesian mixture model that allows for the estimation of clusters
of points characterized by heterogeneous ID. In this article, we focus
our attention on the exposition of TWO-NN and HIDALGO, and we discuss
the pros and cons of both models with the aid of simulated data. More
details about the additional routines, such as GRIDE, are reported in
Section A of the Appendix. Finally, in Section B of the Appendix, we
elaborate more on the strengths and weaknesses of the methods
implemented in \texttt{intRinsic} in comparison to the other existing
packages.

Broadly speaking, the package contains two sets of functions, organized
into high-level and low-level routines. The former set contains
user-friendly and straightforward \texttt{R} functions. Our goal is to
make the package as accessible and intuitive as possible by automating
most tasks. The low-level routines are not exported, as they represent
the package's core. The most computationally-intensive low-level
functions are written in \texttt{C++}, exploiting the interface with
\texttt{R} provided by the packages \texttt{Rcpp} and \texttt{RcppArmadillo}
\citep{Eddelbuettel2011,armad}. The \texttt{C++} implementation
considerably speeds up time-consuming tasks, like running the Gibbs
sampler for the Bayesian mixture model HIDALGO. Moreover,
\texttt{intRinsic} is well integrated with external \texttt{R} packages.
For example, we enriched the package's functionalities defining ad-hoc
methods for generic functions like \texttt{autoplot()} from the
\texttt{ggplot2} package \citep{GGPLOT} to produce the detailed graphical
outputs.

The article is structured as follows.
Section\nobreakspace{}\ref{sec:mod} introduces and describes the
theoretical background of the TWO-NN and HIDALGO methods.
Section\nobreakspace{}\ref{sec:examples} illustrates the basic usage of
the implemented routines on simulated data. We show how to obtain,
manipulate, and interpret the different outputs. Additionally, we assess
the robustness of the methods by monitoring how the results vary when
the input parameters change. Section\nobreakspace{}\ref{sec:alon}
presents an application to a famous real microarray dataset. Finally,
Section\nobreakspace{}\ref{sec:summary} concludes by discussing future
directions and potential extensions to the package.

\section{The modeling background}
\label{sec:mod}

Let \(\bm{X}\) be a dataset with \(n\) data points measured over \(D\)
variables. We denote each observation as \(x_i\in \mathbb{R}^D\), with
\(i=1,\ldots,n\). Despite being observed over a \(D\)-dimensional space,
we suppose that the points take place on a latent manifold
\(\mathcal{M}\) with intrinsic dimension \(d\leq D\). Generally, we
expect that \(d <<D\). Thus, we postulate that a low-dimensional
data-generating mechanism can accurately describe the dataset.\\
Then, consider a single data point \(x_i\). Starting from this point,
one can order the remaining \(n-1\) observations according to their
distance from \(x_i\). This way, we obtain a list of nearest neighbors
(NNs) of increasing order. Formally, let
\(\Delta:\mathbb{R}^D\times\mathbb{R}^D\rightarrow\mathbb{R}^+\) be a
generic distance function between data points. We denote with
\(x_i^{(l)}\) the \(l\)-th NN of \(x_i\) and with
\(r_{i,l}=\Delta\left(x_{i},x_i^{(l)}\right)\) their distance, for
\(l=1,\ldots, n-1\). Given the sequence of NNs for each data point, we
can define the
\emph{volume of the hyper-spherical shell enclosed between two successive neighbors of}
\(x_i\) as \begin{equation}
	\nu_{i,l}=\omega_{d}\left(r_{i,l}^{d}-r_{i,l-1}^{d}\right), \quad \quad \text{for  }l=1,\ldots,n-1,\;\text{ and }\;i=1,\ldots,n, \label{eq::HSshell}
\end{equation} where \(d\) is the dimensionality of the latent manifold
in which the points are embedded (the ID) and \(\omega_{d}\) is the
volume of the \(d\)-dimensional hyper-sphere with unitary radius. For
this formula to hold, we need to set \(x_{i,0}\equiv x_{i}\) and
\(r_{i,0}=0\). We provide a visual representation of the introduced
quantities in Figure\nobreakspace{}\ref{fig:concentric} for \(l=1,2\),
which depicts the three-dimensional case.

From a modeling perspective, we assume that the dataset \(\bm{X}\) is a
realization of a Poisson point process characterized by density function
\(\rho\left(x\right)\). \citet{Facco} showed that the hyper-spherical
shells defined in Equation\nobreakspace{}\ref{eq::HSshell} are the
multivariate extension of the well-known \emph{inter-arrival times}
\citep{Kingman1992}. Indeed, they proved that under the assumption of
homogeneity of the Poisson point process, i.e.,
\(\rho(x)=\rho\:\: \forall x\), all the \(\nu_{i,l}\)'s are
independently drawn from an Exponential distribution with rate equal to
the density \(\rho\): \(\nu_{i,l}\sim Exp(\rho)\), for
\(l =1,\ldots,n-1,\) and \(i=1,\dots,n\). This fundamental result
motivates the derivation of the estimators we will introduce in the
following.

\begin{figure}[t!]
	\begin{center}
\resizebox{8cm}{8cm}{%
	\begin{tikzpicture}
\tikzstyle{every node}=[font=\LARGE]

\shade[ball color = blue!40, opacity = 0.3] (0,0) circle (8cm);
\shade[ball color = blue!40, opacity = 0.3] (0,0) circle (6cm);
\shade[ball color = blue!40, opacity = 0.3] (0,0) circle (3cm);
\draw (0,0) circle (8cm) node[below left] {$\bm{x}_i$};
\draw[loosely dotted] (0,0) circle (6cm);
\draw[loosely dotted] (0,0) circle (3cm);
\draw[loosely dotted] (-8,0) arc (180:360:8 and 2.6);
\draw[loosely dotted] (-6,0) arc (180:360:6 and 1.6);
\draw[loosely dotted] (-3,0) arc (180:360:3 and 0.6);
\draw[loosely dotted] (8,0) arc (0:180:8 and 2.6);
\draw[loosely dotted] (6,0) arc (0:180:6 and 1.6);
\draw[loosely dotted] (3,0) arc (0:180:3 and 0.6);

\fill[fill=black] (0,0) circle (4pt);
\fill[fill=black] (-2, 2.2360679775) circle (4pt) node[above left] {$\bm{x}_{(i,1)}$};
\fill[fill=black] (4, 4.472135955) circle (4pt) node[above right] {$\bm{x}_{(i,2)}$};
\fill[fill=black] (2, -7.74596669241) circle (4pt) node[below right = .2cm] {$\bm{x}_{(i,3)}$};

\draw[dashed] (0,0 ) -- node[above,left]{$r_{i,1}$} (-2, 2.2360679775)  
;
\draw[dashed] (0,0 ) -- node[above,left]{$r_{i,2}$} (4, 4.472135955);
\draw[dashed] (0,0 ) -- node[above,right]{$r_{i,3}$} (2, -7.74596669241);
	\end{tikzpicture}
}
	\end{center}
	\caption{An illustration in three dimensions of the quantities involved in the TWO-NN modeling framework. The central dot represents the $i$-th data point. The selected observation is connected by dashed lines, representing the distances $r_{i,j}$, $j=1,2,3$, to its first three NNs. The different spherical shells have volumes $v_{i,j}$, $j=1,2,3$.}  
	\label{fig:concentric}
\end{figure}

\subsection{The TWO-NN estimator}
\label{sec:twonn}

Building on the distribution of the hyper-spherical shells,
\citet{Facco} noticed that if the intensity of the underlying Poisson
point process is assumed to be constant on the scale of the second NN,
the following distributional result holds: \begin{equation}
	\mu_{i,1,2} = \frac{r_{i,2}}{r_{i,1}} \sim Pareto(1,d), \quad \quad \mu_i \in \left[1,+\infty\right) \quad \quad i=1,\ldots,n.
	\label{eq::firstresult}    
\end{equation} In other words, if the intensity of the Poisson point
process that generates the data can be regarded as locally constant (on
the scale of the second NN), the ratio of the first two NN distances
from each point is Pareto distributed. Recall that the Pareto random
variable is characterized by a scale parameter \(a\), shape parameter
\(b\), and density function \(f_X(x)=ab^a x^{-a-1}\) defined over
\(x\in\left[a,+\infty\right)\). Remarkably,
Equation\nobreakspace{}\ref{eq::firstresult} states that the ratio
\({r_{i,2}}/{r_{i,1}}\) follows a Pareto distribution with scale \(a=1\)
and shape \(b=d\), i.e., the shape parameter can be interpreted as the
ID of the dataset. Estimating \(d\) by exploiting the ratios of
distances between each point and its first two NNs is the core of the
TWO-NN procedure.

One can also attain more general results by considering ratios of
distances with NNs of generic orders. A generalized ratio will be
denoted with \(\mu_{i,n_1,n_2} = r_{i,n_2}/r_{i,n_1}\) for
\(i = 1,\ldots, n\). Here, \(n_1\) and \(n_2\) are the NN orders,
integer numbers that need to comply with the following constraint:
\(1\leq n_1 < n_2\leq n\). In this paper, we will mainly focus on
methods involving the ratio of the first two NN distances. The
generalized ratios will be only mentioned when discussing the function
\texttt{compute\_mus()}, for the sake of completeness. Therefore, to
simplify the notation, we will write \(\mu_i = \mu_{i,1,2}\) and
\(\bm{\mu}=\left(\mu_i\right)_{i=1}^n\).\\
Once the vector \(\bm{\mu}\) is computed, we can employ different
estimation techniques for the TWO-NN model. All of the following methods
can be called via the \texttt{intRinsic} function \texttt{twonn()}. The
reader can find examples of its usage in
Section\nobreakspace{}\ref{sec:illustration_twonn}.

\textbf{Linear estimator}. \citet{Facco} proposed to estimate the ID via
the linearization of the Pareto c.d.f. \(F({\mu_i})= (1-\mu_i^{-d})\).
The estimate \(\hat{d}_{OLS}\) is obtained as the solution of
\begin{equation}
	-\log(1-\hat{F}(\mu_{(i)}) )= d\log(\mu_{(i)}),
	\label{regression}     
\end{equation} where \(\hat{F}(\cdot)\) denotes the empirical c.d.f. of
the sample and the \(\mu_{(i)}\)'s are the ratios defined in
Equation\nobreakspace{}\ref{eq::firstresult} sorted by increasing order.
The authors also suggested trimming from \(\bm{\mu}\) a percentage
\(c_{TR}\) of the most extreme values to obtain a more robust
estimation. This choice is justified because the extreme ratios often
correspond to observations that do not comply with the local homogeneity
assumption.

\textbf{Maximum likelihood estimator}. In a similar spirit,
\cite{Denti2021} took advantage of the distributional results in
Equation\nobreakspace{}\ref{eq::firstresult} to derive a simple maximum
likelihood estimator (MLE) and the corresponding confidence interval
(CI). Trivially, the (unbiased) MLE for the shape parameter of a Pareto
distribution is given by: \begin{equation}
	\hat{d} = \frac{n-1}{\sum_{i}^n\log(\mu_i)},
	\label{MLE:TWONN}
\end{equation} while the CI of level (1-\(\alpha\)) is defined as\\
\begin{equation}
	CI(d,1-\alpha)=\left[\frac{\hat{d}}{q^{1-\alpha/2}_{IG_{n,(n-1)}}},\frac{\hat{d}}{q^{\alpha/2}_{IG_{n,(n-1)}}}\right],
	\label{MLE:CI}
\end{equation} where \(q^{\alpha/2}_{IG_{a,b}}\) denotes the quantile of
order \(\alpha/2\) of an Inverse-Gamma distribution of shape \(a\) and
scale \(b\).

\textbf{Bayesian estimator}. It is also straightforward to derive an
estimator according to a Bayesian perspective \citep{Denti2021},
obtained with the specification of a prior distribution on the shape
parameter \(d\). The most natural prior to choose is a conjugate
\(d\sim Gamma(a,b)\). It is immediate to derive the posterior
distribution for the shape parameter: \begin{equation}
	d|\bm{\mu} \sim Gamma\left(a+n, b+\sum_{i=1}^n \log(\mu_i)\right).
\end{equation} With this method, one can quickly recover the principal
quantiles of the posterior distribution and obtain point estimates and
uncertainty quantification with a credible interval (CRI) of level
\(\alpha\).

\subsection{HIDALGO: the heterogeneous intrinsic dimension algorithm}
\label{sec:hidalgo}

The TWO-NN model implicitly assumes the existence of a single manifold.
However, postulating a global, unique ID value for the entire dataset
can often be limiting, especially when the data present complex
dependence structures among variables. To extend the previous modeling
framework, one can imagine that the data points are divided into
clusters, each belonging to a latent manifold with its specific ID.
\citet{Allegra} employed this heterogeneous ID estimation approach in
their model: the heterogeneous ID algorithm (HIDALGO). The authors
proposed as density function of the generating point process a mixture
of \(K\) distributions defined on \(K\) different latent manifolds,
expressed as
\(\rho\left(\bm{x}\right) = \sum_{k=1}^{K}\pi_k\rho_k\left(\bm{x}\right)\),
where \(\bm{\pi}=\left(\pi_1,\ldots,\pi_K\right)\) is the vector of
mixture weights. This assumption induces a mixture of Pareto
distributions as the distribution of the ratios \(\mu_i\)'s:
\begin{equation}
	f(\mu_i|\bm{d},\bm{\pi}) = \sum_{k=1}^{K}\pi_k \: d_k \mu_i^{-(d_k+1)}, \quad \quad i=1,\ldots,n,
	\label{hid}
\end{equation} where \(\bm{d}=\left(d_1,\ldots,d_K\right)\) is the
vector of ID parameters. \cite{Allegra} adopted a Bayesian perspective,
specifying independent Gamma priors for each element of \(\bm{d}:\)
\(d_k\sim Gamma(a_{d},b_{d})\), and a Dirichlet prior for the mixture
weights \(\bm{\pi}\sim Dirichlet(\alpha_1,\ldots, \alpha_K)\). Regarding
the latter, we suggest setting \(\alpha_1=\ldots=\alpha_K=\alpha<0.05\),
fitting a sparse mixture model as indicated by
\citet{Malsiner-Walli2017}. This prior specification encourages the data
to populate only the necessary number of mixture components. Thus, we
see \(K\) as an upper bound on the number of active clusters
\(K^*\leq K\).

\begin{figure}[t!]
	\centering
	\includegraphics[width=\linewidth]{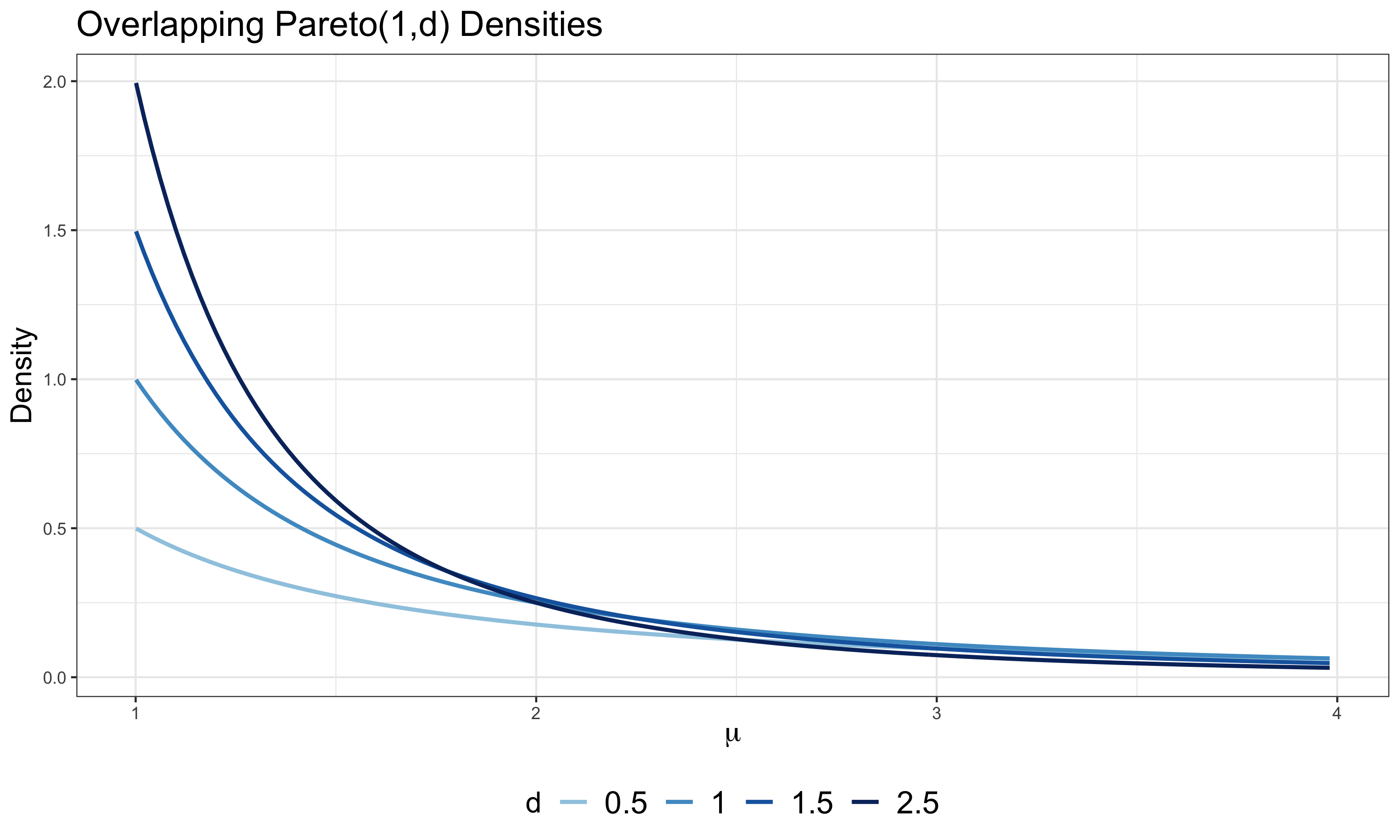}
	\caption{Density functions of Pareto distributions characterized by different shape parameters.}
	\label{fig:pareto}
\end{figure}

Unfortunately, a model-based clustering approach like the one presented
in Equation\nobreakspace{}\ref{hid} is ineffective at modeling the
distance ratios. The problem lies in the fact that the different Pareto
kernels have extremely similar shapes. Therefore, Pareto densities with
varying shape parameters can fit the same data points equally well,
compromising the clustering allocation and the consequent ID estimation.
Even when considering very diverse shape parameters, the right tails of
the resulting Pareto distributions overlap to a great extent. This issue
is evident in Figure\nobreakspace{}\ref{fig:pareto}, where we depict
examples of various \(Pareto(1,d)\) densities.

To address this problem, \citet{Allegra} introduced a local homogeneity
assumption, assuming that neighboring points are more likely to be part
of the same latent manifold. To incorporate this idea in the model, the
authors added an extra penalizing term in the likelihood. We now
summarize their approach.\\
First, they introduced the latent membership labels
\(\bm{z}=(z_1,\ldots,z_n)\) to assign each observation to a cluster,
where \(z_i=k\) means that the \(i\)-th observation belongs to the
\(k\)-th mixture component. Then, they defined the binary adjacency
matrix \(\mathcal{N}^{(q)}\), whose entries are
\(\mathcal{N}_{i,j}^{(q)}=1\) if the point \(x_j\) is among the first
\(q\) NNs of \(x_i\), and 0 otherwise. Finally, they assumed the
following probabilities:
\(\mathbb{P}\left[\mathcal{N}_{i,j}^{(q)}=1|z_i=z_j\right]=\zeta_1\),
with \(\zeta_1>0.5\) and
\(\mathbb{P}\left[\mathcal{N}_{i,j}^{(q)}=1|z_i\neq z_j\right]=\zeta_0\),
with \(\zeta_0<0.5\). These probabilities are employed to define a
distribution over the neighboring structure of each data point \(x_i\):
\(\pi(\mathcal{N}_{i}^{(q)}|\bm{z})=\prod_{j=1}^n \zeta_0^{\mathds{1}_{z_i\neq z_j}}\zeta_1^{\mathds{1}_{z_i=z_j}}/\mathcal{Z}_i\),
where \(\mathcal{Z}_i\) is the normalizing constant and
\(\mathds{1}_{A}\) is the indicator function, equal to 1 when the event
\(A\) is true, 0 otherwise. A more technical discussion of this model
extension and the validity of the underlying hypotheses can be found in
the Supplementary Material of \citet{Allegra}. For simplicity, we assume
\(\zeta_0=\zeta\) and \(\zeta_1= 1-\zeta\). The new likelihood becomes
\begin{equation}
	\mathcal{L}\left(\mu_i,\mathcal{N}^{(q)}|\bm{d},\bm{z},\zeta\right) =  \: d_{z_i} \mu_i^{-(d_{z_i}+1)}\times \prod_{i=1}^{n}\frac{ \zeta^{\mathds{1}_{z_i\neq z_j}}(1-\zeta)^{\mathds{1}_{z_i=z_j}}}{\mathcal{Z}_i}, \quad \quad z_i|\bm{\pi} \sim Cat_{K}(\bm{\pi}),
	\label{MODpara}
\end{equation} where \(Cat_{K}\) denotes a Categorical distribution over
the set \(\{1,\ldots,K\}\). A closed-form for the posterior distribution
is not available, so we rely on MCMC techniques to simulate a posterior
sample.

In our package, this Bayesian mixture model is implemented by the
function \texttt{Hidalgo()}. For the ID parameters, we use conjugate Gamma
prior specifications or variations thereof to account for modeling
inconsistencies. For example, when the nominal dimension \(D\) is low,
the unbounded support of a Gamma prior may provide unrealistic results,
where the posterior distribution assigns positive density to the
interval \(\left(D,+\infty\right)\). \citet{Santos-Fernandez2020}
proposed to employ a more informative prior for \(\bm{d}\):
\begin{equation}
	\pi(d_k) = \hat{\rho}\cdot d_k^{a-1} \exp^{-b d_k} \frac{\mathbbm{1}_{(0,D)}}{\mathcal{C}_{a,b,D}} + (1-\hat\rho)\cdot \delta_D(d_k)  \quad \forall k,
	\label{mixd}
\end{equation} where they denoted the normalizing constant of a
\(Gamma(a,b)\) truncated over \(\left(0,D\right]\) with
\(\mathcal{C}_{a,b,D}\). That is, the prior distribution for \(d_k\) is
a mixture between a truncated Gamma distribution over
\(\left(0,D\right]\) and a point mass located at \(D\). The parameter
\(\rho\) denotes the mixing proportion. When \(\rho=1\), the
distribution in Equation\nobreakspace{}\ref{mixd} reduces to a simple
truncated Gamma. Both approaches are implemented in \texttt{intRinsic}, but
we recommend using the latter. We report the details of the implemented
Gibbs sampler in Section C of the Appendix, while in Section F we
comment more on the concept of \emph{heterogeneous} ID estimation
related to global and local ID definitions.

\section{Examples using {intRinsic}}
\label{sec:examples}

\label{sec:applicat}

This section illustrates the main routines of the \texttt{intRinsic}
package. Here, we indicate the number of observations and the observed
nominal dimension with \texttt{n} and \texttt{D}, respectively. Also,
\(\mathcal{N}_k(m,\Sigma)\) represents a multivariate normal
distribution of dimension \(k\), mean \(m\), and covariance matrix
\(\Sigma\). Moreover, let \(\mathcal{U}^{(k)}_{\left(a,b\right)}\)
represent a multivariate Uniform distribution with support
\(\left(a,b\right)\) in \(k\) dimensions. Finally, we denote with
\(\mathbb{I}_k\) an identity matrix of dimension \(k\).\\
To start, we load our package by running:

\begin{verbatim}
R> library("intRinsic")
\end{verbatim}

\subsection{Simulated datasets}

To illustrate how to apply the different ID estimation techniques
available in the package, we will use three simulated datasets:
Swissroll, Hypercube, and GaussMix. This way, we can compare the results
of the experiments with the ground truth. One can generate the exact
replica of the three simulated datasets used in this paper by running
the code reported below.

The first dataset, Swissroll, is obtained via the classical Swissroll
transformation \(\mathcal{S}:\mathbb{R}^2\rightarrow\mathbb{R}^3\)
defined as \(\mathcal{S}(x,y)=(x\cos(x),y,x\sin(x))\), where each pair
of points \(\left(x,y\right)\) is sampled from two independent Uniform
distributions on \(\left(0,10\right)\). To simulate such a dataset, we
can use the \texttt{intRinsic} function \texttt{Swissroll()}, specifying the
number of observations \texttt{n} as the input parameter.

\begin{verbatim}
R> set.seed(123456)
R> Swissroll <- Swissroll(n = 1000)
\end{verbatim}

The second dataset, Hypercube, contains a cloud of 500 points sampled
from \(\mathcal{U}^{(5)}_{\left(0,1\right)}\) embedded in an
eight-dimensional space \(\mathbb{R}^8\). To fill the gap between the
nominal dimension (8) and the ID (5), we add three columns of zeros.

\begin{verbatim}
R> HyperCube  <- cbind(replicate(5, runif(500)), 0, 0, 0)
\end{verbatim}

Lastly, the dataset GaussMix contains 1500 data points generated from
three random variables all contained in a five-dimensional Euclidean
space. More precisely, we consider a bivariate random variable defined
as \(X_1 = (X_0,3X_0)\) where \(X_0\sim \mathcal{N}_1(-5,1)\), and the
three- and five-dimensional variables \begin{equation*}
	X_2\sim \mathcal{N}_3(0,\mathbbm{I}_3), \quad 
	X_3 \sim \mathcal{N}_5(5,\mathbbm{I}_5).
\end{equation*} We embed each of them in a five-dimensional space by
adding the corresponding number of columns of zeros, as we did for the
HyperCube dataset.

\begin{verbatim}
R> x0 <- rnorm(500, mean = -5, sd = 1)
R> x1 <- cbind(x0, 3 * x0, 0, 0, 0)
R> x2 <- cbind(replicate(3, rnorm(500)), 0, 0)
R> x3 <- replicate(5, rnorm(500, 5))
R> GaussMix <- rbind(x1, x2, x3)
R> class_GMix <- rep(c("A", "B", "C"), rep(500, 3))
\end{verbatim}

\begin{figure}[t!]
	\centering
	\includegraphics[height=10cm,width=10cm]{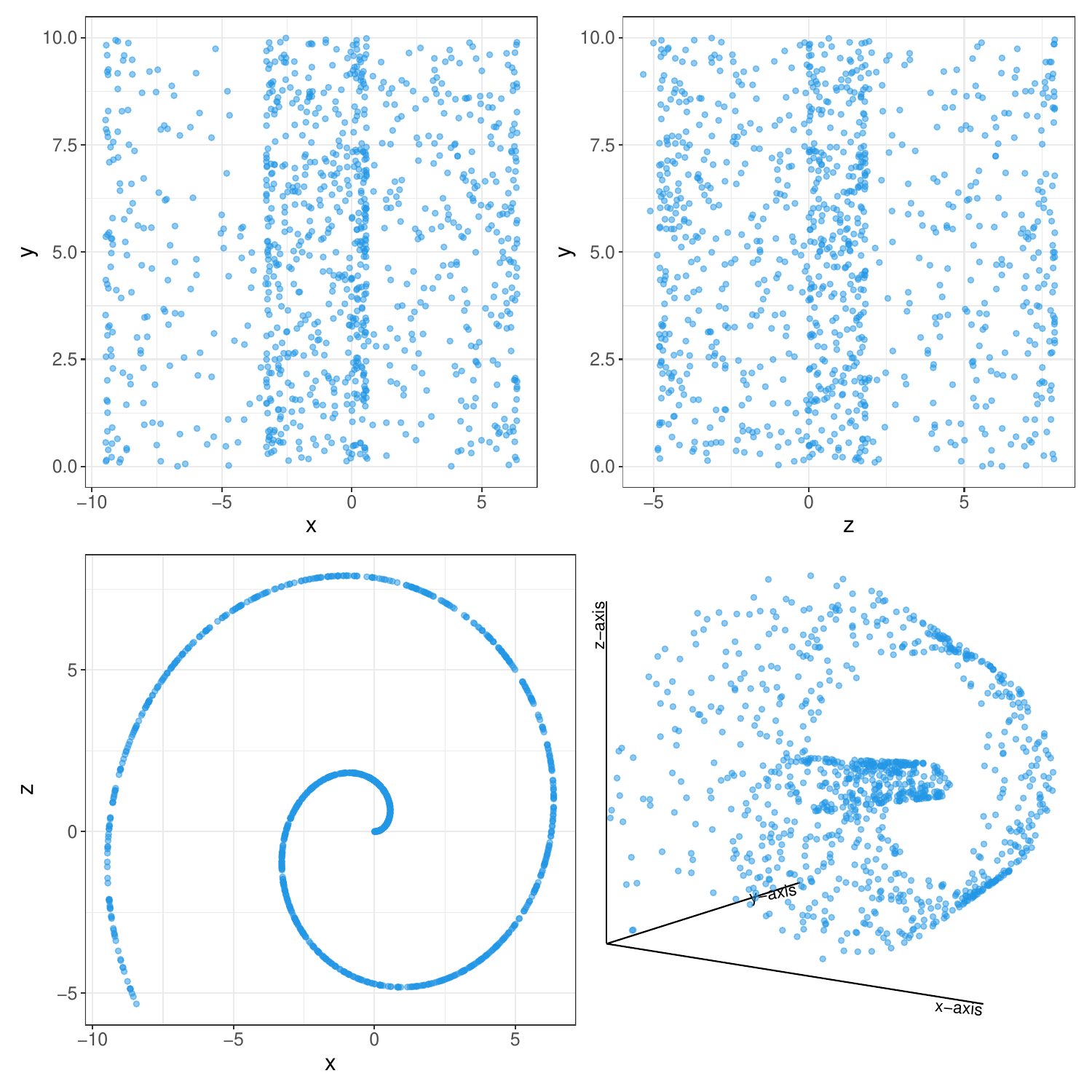}
	\caption{Scatterplots of the three variables in the Swissroll dataset. The dependence among the coordinates $x$ and $z$ is evident.}
	\label{fig:scattSwiss}
\end{figure}

Next, we need to establish the true values of the ID for the different
datasets. Scatterplots are useful exploratory tools to spot any clear
dependence across the columns of a dataset. For example, if we plot all
the possible two-dimensional scatterplots from the Swissroll dataset, we
obtain Figure\nobreakspace{}\ref{fig:scattSwiss}. The different panels
show that two of the three coordinates are free, and we can recover the
last coordinate as a function of the others. Therefore, the ID of
Swissroll is equal to 2. Moreover, from the description of the data
simulation, it is also evident that the ID is equal to 5 for the
Hypercube dataset. However, it is not as simple to figure out the value
of the true ID for GaussMix, given the heterogeneity of its
data-generating mechanism. Table\nobreakspace{}\ref{tab:data} summarizes
the sample sizes along with the true nominal dimensions \texttt{D} and
IDs that characterize the three datasets.

\begin{table}[t!]
	\centering
	\begin{tabular}{lccc}
\toprule
Name & \texttt{n} & \texttt{D} & ID\\
\midrule
Swissroll & 1000  & 3 & 2 \\
Hypercube & 500   & 8 & 5 \\
GaussMix  & 1500  & 5 & ? \\
\bottomrule
	\end{tabular}
	\caption{Summary of the characteristics of the three simulated datasets.}
	\label{tab:data}
\end{table}

\subsection{Ratios of nearest neighbors distances}
\label{sec:computemus}

The ratios of NN distances constitute the core quantities on which the
theoretical development presented in Section\nobreakspace{}\ref{sec:mod}
is based. We can compute the ratios \(\mu_i\) defined in
Equations\nobreakspace{}\ref{eq::firstresult} with the function
\texttt{compute\_mus()}. All in all, the function can also compute the
generalized ratios \(\bm{\mu}_{n_1,n_2}\), where \texttt{n1 < n2}, as
presented in Section\nobreakspace{}\ref{sec:twonn}. In fact, the
function needs the following arguments:

\begin{itemize}
	\item \texttt{X}: a dataset of dimension \texttt{n}$ \times$\texttt{D} of which we want to compute the distance ratios;
	\item \texttt{dist\_mat}: a \texttt{n}$ \times$\texttt{n} symmetric matrix containing the distances between data points which one can pass instead of \texttt{X};
	\item \texttt{n1} and \texttt{n2}: the orders of the closest and furthest nearest neighbors to consider, respectively. As default, \texttt{n1 = 1} and \texttt{n2 = 2}.
\end{itemize}

The function has two additional arguments, \texttt{Nq} and \texttt{q},
that we will introduce later in
Section\nobreakspace{}\ref{sec:hidalgo_app} when we will illustrate the
HIDALGO model.\\
Note that the specification of \texttt{dist\_mat} overrides the argument
passed as \texttt{X}. Instead, if the distance matrix \texttt{dist\_mat}
is not specified, \texttt{generate\_mus()} relies on the function
\texttt{get.knn()} from the package \texttt{FNN} \citep{FNN}, which
implements fast NN-search algorithms on the original dataset.

The main output of the function is the vector of ratios
\(\bm{\mu}_{n_1,n_2}\), an object of class \texttt{mus} for which
appropriate \texttt{print()} and \texttt{plot()} methods are defined. To use
the function, we can easily write:

\begin{verbatim}
R> mus_Swissroll <- compute_mus(X = Swissroll)
R> mus_HyperCube <- compute_mus(X = HyperCube)
R> mus_GaussMix <- compute_mus(X = GaussMix)
\end{verbatim}

Calling the function with default arguments produces
\(\bm{\mu}=\bm{\mu}_{1,2}\). To explicitly compute generalized ratios,
we need to specify the NN orders \texttt{n1} and \texttt{n2}. Here, we
report two different examples:

\begin{verbatim}
R> mus_Swissroll_1 <- compute_mus(X = Swissroll, n1 = 5, n2 = 10)
R> mus_HyperCube_1 <- compute_mus(X = HyperCube, n1 = 5, n2 = 10)
R> mus_GaussMix_1 <- compute_mus(X = GaussMix, n1 = 5, n2 = 10)
\end{verbatim}

and

\begin{verbatim}
R> mus_Swissroll_2 <- compute_mus(X = Swissroll, n1 = 10, n2 = 20)
R> mus_HyperCube_2 <- compute_mus(X = HyperCube, n1 = 10, n2 = 20)
R> mus_GaussMix_2 <- compute_mus(X = GaussMix, n1 = 10, n2 = 20)
R> mus_GaussMix_2
\end{verbatim}
\begin{verbatim}
Ratio statistics mu's:
NN orders: n1 = 10, n2 = 20.
Sample size: 1500.
Nominal Dimension: 5.
\end{verbatim}

\begin{figure}[t!]
	\centering
	\includegraphics[width=.95\linewidth]{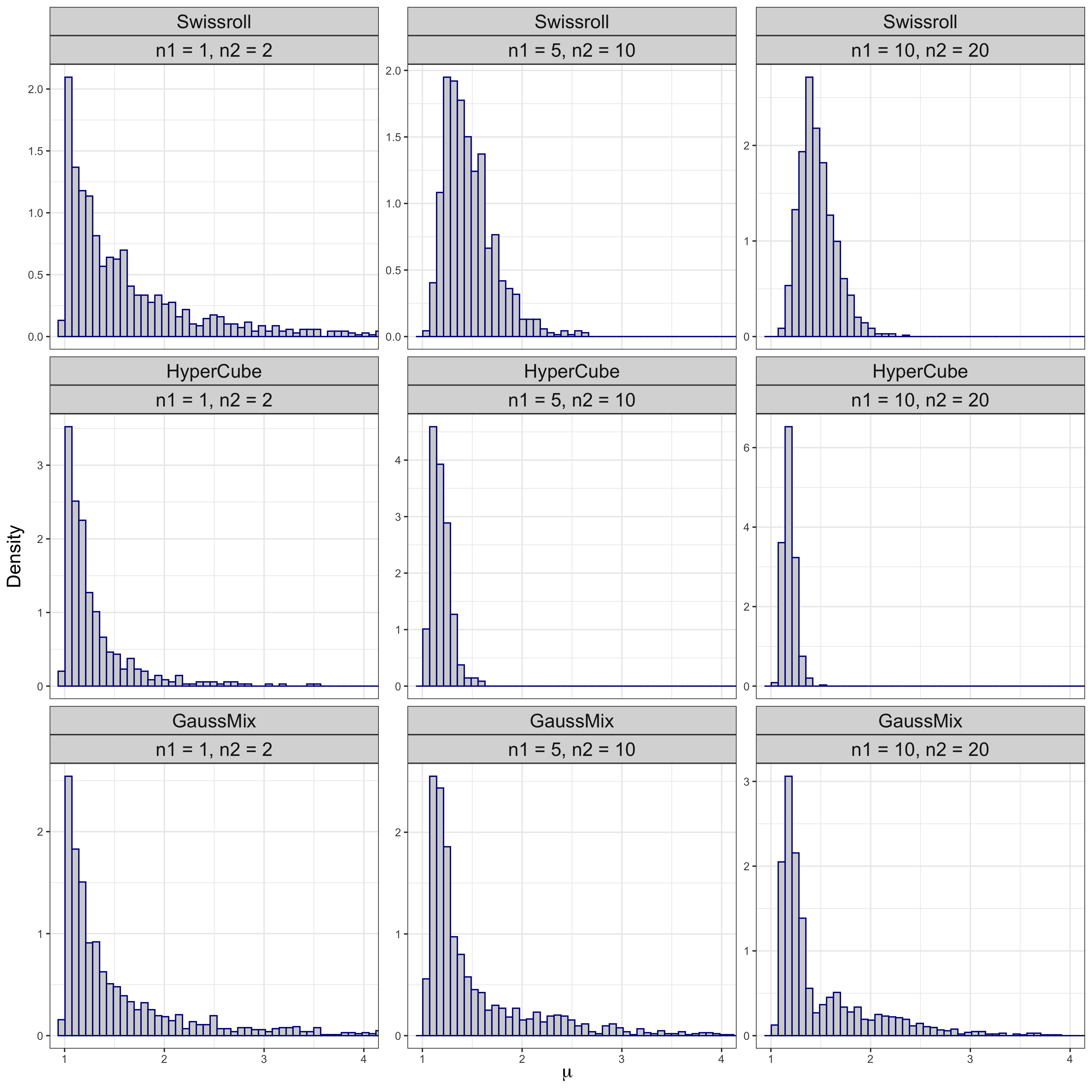}
	\caption{Histograms of the ratios $\bm{\mu}_{n_1,n_2}$ for the Swissroll, Hypercube, and GaussMix datasets. The shape of the histograms in the first column suggests that a Pareto distribution could be a good fit, according to the TWO-NN model.}
	\label{fig:mu_hists}
\end{figure}

The histograms of the computed ratios are presented in
Figure\nobreakspace{}\ref{fig:mu_hists}. The panels in each row
correspond to different datasets, while the varying NN orders are
reported by column. The horizontal axes are truncated over the interval
\(\left[0,4\right]\) to improve the visualization. The histograms
present the right-skewed shape typical of the Pareto distribution.
However, some ratios could assume extreme values, especially when low
values of NN orders are chosen. To provide an example, in
Table\nobreakspace{}\ref{tab:summary_gmx} we display the summary
statistics of the three vectors of ratios computed on GaussMix. The
maximum in the first line has a high magnitude, but it significantly
reduces when higher NN orders are considered. It is also interesting to
observe how the distribution for the ratios of GaussMix when
\(\texttt{n1} = 10\) and \(\texttt{n2} = 20\) (bottom-right panel) is
multimodal, a symptom of the presence of heterogeneous manifolds. We
remark again that we will focus on the TWO-NN and HIDALGO models for the
rest of the paper. Therefore, we will only use \texttt{compute\_mus()} in
its default specification, simply computing
\(\bm{\mu} = \left(\mu_i\right)_{i=1}^n\). The ratios of NNs of generic
order are necessary when using the GRIDE model. See Section A of the
Appendix and \cite{Denti2021} for more details.

\begin{table}[th!]
	\centering
	\begin{tabular}{cccccccc}
\toprule
\texttt{n1} & \texttt{n2} & Minimum & 1st quartile & Median & Mean & 3rd quartile & Maximum\\
\midrule
1  & 2   & 1.0002 & 1.1032 & 1.3048 & 18.4239 & 1.9228 & 5874.3666\\
5  & 10  & 1.0234 & 1.1627 & 1.2858 & 1.5344  & 1.7043 & 6.9533   \\
10 & 20  & 1.0472 & 1.1685 & 1.2613 & 1.4987  & 1.7250 & 4.0069   \\
\bottomrule
	\end{tabular}
	\caption{Summary statistics of the generalized ratios obtained from the GaussMix dataset. Each row corresponds to a different combination of NN orders.}
	\label{tab:summary_gmx}
\end{table}

Finally, recall that the model is based on the assumption that a Poisson
point process is the generating mechanism of the dataset. Ergo, the
model cannot handle ties among data points. From a more practical point
of view, if \(\exists i\neq j\) such that \(x_i=x_j\), the computation
of \(\mu_i\) would be unfeasible since \(r_{i,1}=0\). We devised the
function \texttt{compute\_mus()} to automatically detect if duplicates are
present in a dataset. In that case, the function removes the duplicates
and provides a warning. We showcase this behavior with a simple example:

\begin{verbatim}
R> Dummy_Data_with_replicates <- rbind(
+     c(1, 2, 3), c(1, 2, 3),
+     c(1, 4, 3), c(1, 4, 3),
+     c(1, 4, 5) )
R> mus_shorter <- compute_mus(X = Dummy_Data_with_replicates)
\end{verbatim}
\begin{verbatim}
Warning: 
Duplicates are present and will be removed.

Original sample size: 5. New sample size: 3.
\end{verbatim}

The function \texttt{compute\_mus()} is at the core of many other
high-level routines we use to estimate the ID. The following subsection
shows how to implement the TWO-NN model to obtain a point estimate of a
global, homogeneous ID accompanied by the corresponding CIs or CRIs.

\subsection{Estimating a global ID value with TWO-NN}
\label{sec:illustration_twonn}

We showcase how to carry out inference on the ID with the TWO-NN model
using linear, MLE, and Bayesian estimation methods. The low-level
functions that implement these methods are \texttt{twonn\_mle()},
\texttt{twonn\_linfit()}, and \texttt{twonn\_bayes()}, respectively. One can
call these low-level functions via the high-level function
\texttt{twonn()}. Regardless of the preferred estimation method, the
\texttt{twonn()} function takes the following arguments: the dataset
\texttt{X} or the distance matrix \texttt{dist\_mat} (refer to previous
input descriptions for more details), along with

\begin{itemize}
	\item \texttt{mus}: the vector of second-to-first NN distance ratios. If this argument is provided, \texttt{X} and \texttt{dist\_mat} will be ignored;
	\item \texttt{method}: a string stating the preferred estimation method. Could be \texttt{"mle"} (the default), \texttt{"linfit"}, or \texttt{"bayes"};
	\item \texttt{alpha}: the confidence level (for \texttt{"mle"} and \texttt{"linfit"}) or the posterior probability included in the CRI (\texttt{"bayes"});
	\item \texttt{c\_trimmed}: the proportions of most extreme ratios to exclude from the analysis.
\end{itemize}

The object that the function returns is a list,characterized by a class
that varies according to the selected estimation method. Tailored
\texttt{R} methods have been devised to extend the generic functions
\texttt{print()}, \texttt{summary()}, \texttt{plot()}, and \texttt{autoplot()}
to interact with these new classes. The first element of the returned
list always contains the estimates, while the others provide additional
information about the chosen estimation process.

\textbf{Linear estimator}. We apply the linear estimator to the
Swissroll dataset. As an example, we fit five linear models by setting
\texttt{method = "linfit"}, adopting different trimming proportions. The
function \texttt{summary()} provides an informative recap of the
estimation process. We show the results for \texttt{lin\_2} as an
example.

\begin{verbatim}
R> lin_1 <- twonn(X = Swissroll, method = "linfit", c_trimmed = 0)
R> lin_2 <- twonn(X = Swissroll, method = "linfit", c_trimmed = 0.001)
R> lin_3 <- twonn(X = Swissroll, method = "linfit", c_trimmed = 0.01)
R> lin_4 <- twonn(X = Swissroll, method = "linfit", c_trimmed = 0.05)
R> lin_5 <- twonn(X = Swissroll, method = "linfit", c_trimmed = 0.1)
R> summary(lin_2)
\end{verbatim}
\begin{verbatim}
Model: TWO-NN
Method: Least Squares Estimation
Sample size: 1000, Obs. used: 999. Trimming proportion: 0.1%
ID estimates (confidence level: 0.95)

| Lower Bound| Estimate| Upper Bound|
|-----------:|--------:|-----------:|
|    1.986659| 1.999682|    2.012705|
\end{verbatim}

The results of these experiments are collected in
Table\nobreakspace{}\ref{tab:res_lin}. This first example allows us to
comment on the trimming level to choose. Trimming the most extreme
observations may be fundamental since outliers may distort the estimate.
However, too much trimming would remove important information regarding
the tail of the Pareto distribution, which is essential for the correct
estimation of the ID. The estimates improve for very low levels of
trimming but start to degenerate as more than 5\% of observations are
removed from the dataset.

\begin{table}[t!]
	\centering
	\begin{tabular}{lcccccc}
\toprule
Trimming percentage & 0\% & 0.1\% & 1\% & 5\% & 10\%\\
\midrule
Lower bound  & 1.9524 & 1.9867 & 2.2333 & 2.5709 & 2.9542\\
Estimate  & 1.9669 & 1.9997 & 2.2457 & 2.5988 & 2.9974\\
Upper bound  & 1.9814 & 2.0127 & 2.2581 & 2.6267 & 3.0407\\
\bottomrule
	\end{tabular}
	\caption{Point estimates and relative CIs for the ID values retrieved from the Swissroll dataset with the linear estimator. Each column displays the results for a specific trimming level.}
	\label{tab:res_lin}
\end{table}

We can also visually assess the goodness of fit via a dedicated
\texttt{autoplot()} function, which displays the data and the estimated
regression line. The slope of the regression lines corresponds to the
linear fit ID estimates. For example, we obtain the plots in
Figure\nobreakspace{}\ref{fig:lin_fit_trim} with the following two lines
of code:

\begin{verbatim}
R> autoplot(lin_1, title = "No trimming")
R> autoplot(lin_5, title = "10% trimming")
\end{verbatim}

\begin{figure}[th!]
	\centering
	\includegraphics[width=\linewidth]{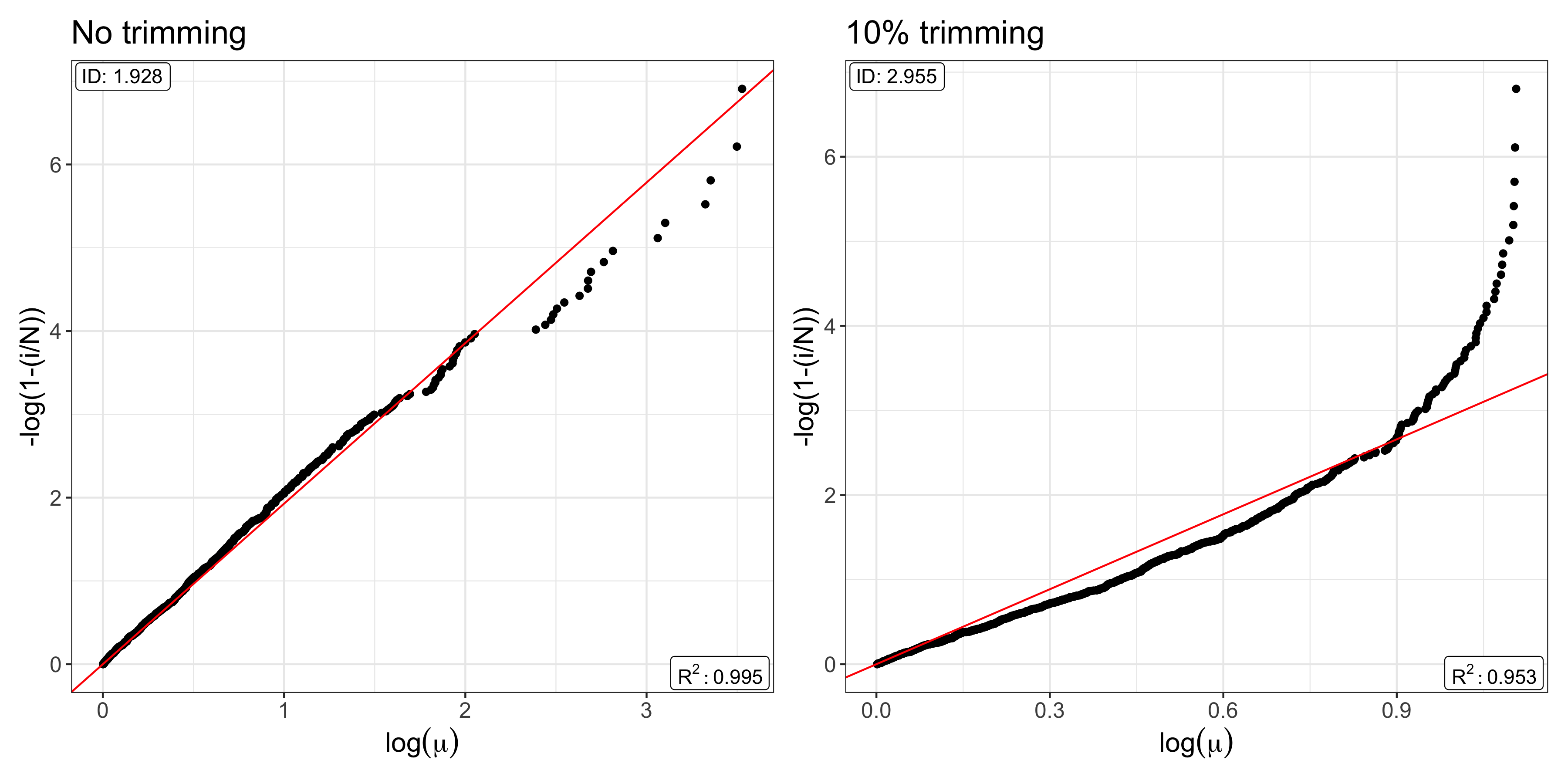}
	\caption{Swissroll dataset. Regression lines estimated from $ -\log(1-\hat{F}(\mu_i) )= d\log(\mu_i)$ with no trimming (left panel) and 10\% of trimmed observations (right panel).}
	\label{fig:lin_fit_trim}
\end{figure}

\textbf{MLE.} A second way to obtain an ID estimate, along with its CI,
is via MLE. The formulas implemented are presented in
Equations\nobreakspace{}\ref{MLE:TWONN} and\nobreakspace{}\ref{MLE:CI}.
We compute the MLE by calling the low-level \texttt{twonn\_mle()} function
setting \texttt{method = "mle"}. In addition to the previous arguments,
one can also specify

\begin{itemize}
	\item \texttt{unbiased}: logical, if \texttt{TRUE} the point estimate according to the unbiased estimator (where the numerator is $n-1$, as in Equation\nobreakspace{}\ref{MLE:TWONN}) is computed.
\end{itemize}

We compute the ID on Hypercube via MLE using different distance
definitions: Euclidean, Manhattan, and Canberra. These distances are
calculated with the \texttt{dist()} function of the \texttt{stats} package.

\begin{verbatim}
R> dist_Eucl_D2 <- dist(HyperCube)
R> dist_Manh_D2 <- dist(HyperCube, method = "manhattan")
R> dist_Canb_D2 <- dist(HyperCube, method = "canberra")
\end{verbatim}

Other distance matrices can be employed as well. In this example, we
also show how the widths of the CIs change by varying the confidence
levels. We print the results stored in the object \texttt{mle\_12} as an
example. We write:

\begin{verbatim}
R> mle_11 <- twonn(dist_mat = dist_Eucl_D2)
R> mle_12 <- twonn(dist_mat = dist_Eucl_D2, alpha = .99)
R> mle_21 <- twonn(dist_mat = dist_Manh_D2)
R> mle_22 <- twonn(dist_mat = dist_Manh_D2, alpha = .99)
R> mle_31 <- twonn(dist_mat = dist_Canb_D2)
R> mle_32 <- twonn(dist_mat = dist_Canb_D2, alpha = .99)
R> summary(mle_12)
\end{verbatim}
\begin{verbatim}
Model: TWO-NN
Method: MLE
Sample size: 500, Obs. used: 495. Trimming proportion: 1%
ID estimates (confidence level: 0.99)

| Lower Bound| Estimate| Upper Bound|
|-----------:|--------:|-----------:|
|    3.968082| 4.459427|     5.00273|
\end{verbatim}

The results are reported in Table\nobreakspace{}\ref{tab:MLEtab}. The
type of distance can lead to differences in the results. Overall, the
estimators agree with each other, obtaining values close to the ground
truth.

\begin{table}[th!]
	\centering
	\begin{tabular}{lcccccc}
\toprule
\texttt{dist()} & \multicolumn{2}{c}{\texttt{Euclidean}} & \multicolumn{2}{c}{\texttt{Manhattan}}& \multicolumn{2}{c}{\texttt{Canberra}}\\ 
$\alpha$ &  $0.95$ & $0.99$ & $0.95$ &  $0.99$ & $0.95$ & $0.99$\\
\midrule
Lower bound & 4.0834 & 3.9681 & 4.0736 & 3.9585 & 4.6001 & 4.4701\\
Estimate & 4.4594 & 4.4594 & 4.4487 & 4.4487 & 5.0237 & 5.0237\\
Upper bound & 4.8706 & 5.0027 & 4.8588 & 4.9906 & 5.4868 & 5.6357\\
\bottomrule
	\end{tabular}
	\caption{MLEs obtained with the TWO-NN model applied to the Hypercube dataset. Different distance functions and confidence level specifications are adopted.}
	\label{tab:MLEtab}
\end{table}

\textbf{Bayesian estimator.} The third option for ID estimation is to
adopt a Bayesian perspective and specify a prior distribution for the
parameter \(d\). To obtain the Bayesian estimates, we call the low-level
function \texttt{twonn\_bayes()} setting \texttt{method = "bayes"} in
\texttt{twonn()}. Along with the arguments mentioned above, we can also
specify the following:

\begin{itemize}
	\item  \texttt{a\_d} and \texttt{b\_d}: shape and rate parameters of the Gamma prior distribution on $d$. A vague specification is adopted as default with  \texttt{a\_d = 0.001} and \texttt{b\_d = 0.001}. This implies $\mathbb{E}(d) = 1,$ and $\mathbb{V}ar(d) = 1000$.
\end{itemize}

Differently from the previous two cases, \texttt{alpha} is now assumed
to be the probability contained in the CRI computed from the posterior
distribution. Along with the CRI, the function outputs the posterior
mean, median, and mode. In the following, four examples showcase the
usage of this function on the Swissroll dataset with different
combinations of \texttt{alpha} and Gamma hyperparameters. The results
are summarized in Table\nobreakspace{}\ref{tab:baytownn}.

\begin{verbatim}
R> bay_1 <- twonn(X = Swissroll, method = "bayes")
R> bay_2 <- twonn(X = Swissroll, method = "bayes", alpha = 0.99)
R> bay_3 <- twonn(X = Swissroll, method = "bayes", a_d = 1, b_d = 1)
R> bay_4 <- twonn(X = Swissroll, method = "bayes", a_d = 1, b_d = 1, alpha = 0.99)
\end{verbatim}

\begin{figure}[t!]
	\centering
	\includegraphics[width=\linewidth]{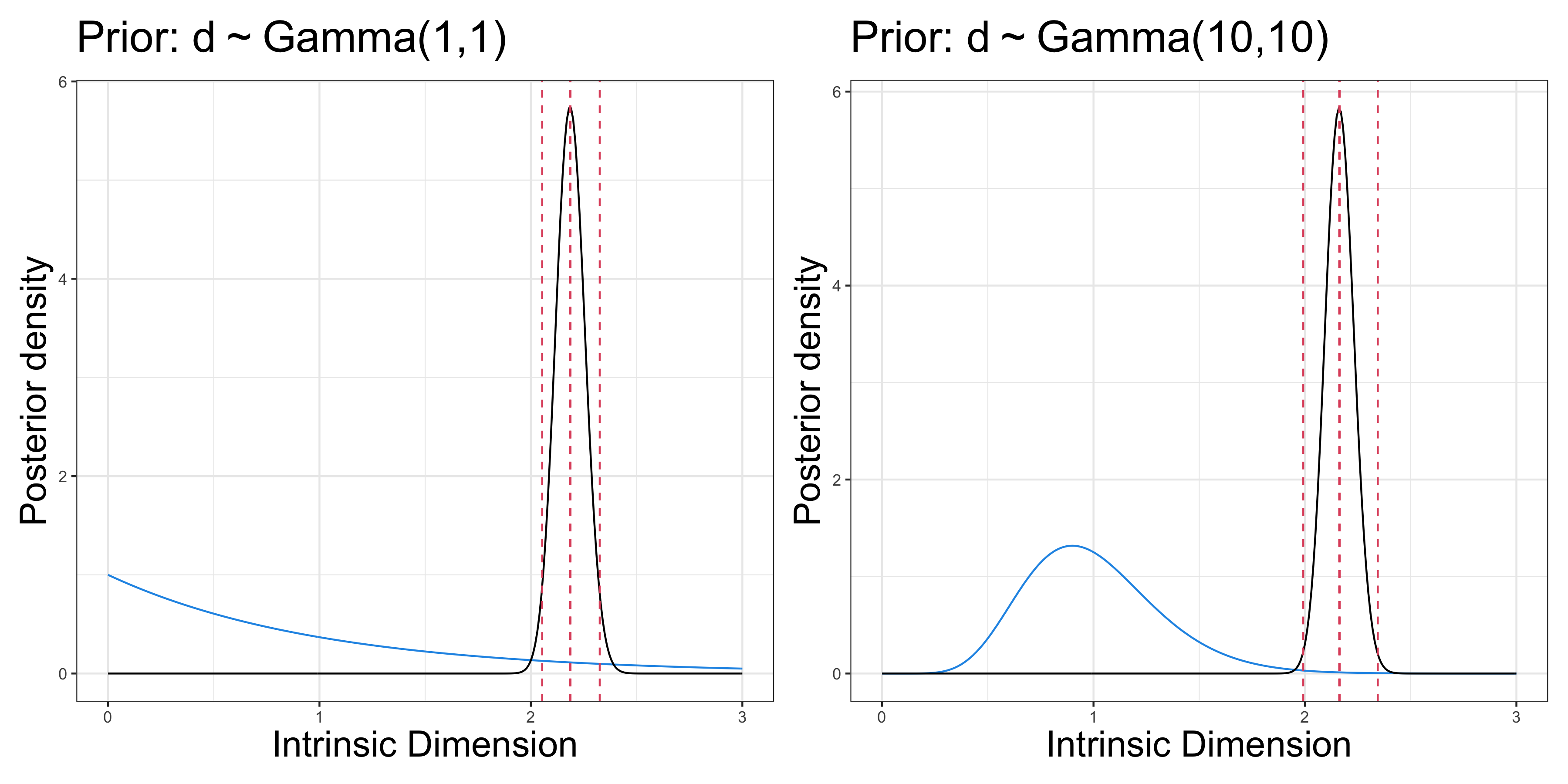}
	\caption{Swissroll dataset. Graphical representation of the posterior distribution (black line), prior distribution (blue line), and main quantiles and average (vertical dotted red lines) under $d\sim Gamma(1,1)$ (left panel) and $d\sim Gamma(10,10)$ (right panel) prior specifications.}
	\label{fig:bayes}
\end{figure}

We can plot the posterior density of the parameter \(d\) using
\texttt{autoplot()}, as displayed in Figure\nobreakspace{}\ref{fig:bayes}.
When plotting an object of class \texttt{twonn\_bayes}, we can also
specify the following parameters:

\begin{itemize}
	\item  \texttt{plot\_low} and \texttt{plot\_upp}:  lower and upper extremes of the support on which the posterior is evaluated;
	\item \texttt{by}: increment of the sequence going from \texttt{plot\_low} to \texttt{plot\_upp} that defines the support. 
\end{itemize}

As an example, we compare the prior specification used for the object
\texttt{bay\_4} (\(d\sim Gamma(1,1)\)) with a more informative one
(\(d\sim Gamma(10,10)\)) by writing:

\begin{verbatim}
R> bay_5 <- twonn(X = Swissroll, method = "bayes", a_d = 10, b_d = 10, 
+                 alpha = 0.99)
R> summary(bay_5)
\end{verbatim}
\begin{verbatim}
Model: TWO-NN
Method: Bayesian Estimation
Sample size: 1000, Obs. used: 990. Trimming proportion: 1%
Prior d ~ Gamma(10, 10)
Credibile Interval quantiles: 0.5%, 99.5%
Posterior ID estimates:

| Lower Bound|     Mean|   Median|     Mode| Upper Bound|
|-----------:|--------:|--------:|--------:|-----------:|
|    1.991901| 2.164113| 2.163391| 2.161949|    2.344453|
\end{verbatim}

The posterior distribution is depicted in black, the prior in blue, and
the dashed vertical red lines represent the estimates.

\begin{table}[th!]
	\centering
	\begin{tabular}{lcccc}
\toprule
Prior  & \multicolumn{2}{c}{Default} & \multicolumn{2}{c}{$d\sim Gamma(1,1)$}\\
$\alpha$ & $0.95$ & $0.99$ & $0.95$ & $0.99$\\
\midrule
Lower bound & 2.0556 & 2.0147 & 2.0532 & 2.0124\\
Mean        & 2.1899 & 2.1899 & 2.1872 & 2.1872\\
Median      & 2.1891 & 2.1891 & 2.1865 & 2.1865\\
Mode        & 2.1876 & 2.1876 & 2.1850 & 2.1850\\
Upper bound & 2.3284 & 2.3733 & 2.3255 & 2.3703\\
\bottomrule
	\end{tabular}
	\caption{Swissroll dataset. Posterior estimates under the Bayesian specification according to different prior specifications and CRI levels $\alpha$.}
	\label{tab:baytownn}
\end{table}

So far, we have discussed methods to determine a global ID estimate
accurately and efficiently. Knowing the simulated data-generating
processes, we could easily compare the obtained estimates with the
ground truth for the Swissroll and Hypercube datasets. However, the same
task is not immediate when dealing with GaussMix. For GaussMix, relying
only on a global ID estimate may constitute an oversimplification since
the data points are generated from Gaussian distributions defined over
different manifolds of heterogeneous dimensions. This scenario is more
likely to occur with datasets describing real phenomena, often
characterized by complex dependencies, and it will be the focus of the
next section. \newpage

\subsection{Detecting manifolds with heterogeneous ID using HIDALGO}
\label{sec:hidalgo_app}
\subsubsection{Detecting the presence of multiple manifolds}

In contexts where data may exhibit heterogeneous ID, we face two main
challenges: (i) detect the actual presence of multiple manifolds in the
data, and (ii) accurately estimate their IDs. To tackle these problems,
we start by applying the \texttt{twonn()} function to GaussMix with
\texttt{method} equal to \texttt{"linfit"} and \texttt{"mle"}.

\begin{verbatim}
R> mus_gm <- compute_mus(GaussMix)
R> summary(twonn(mus = mus_gm, method = "linfit"))
\end{verbatim}
\begin{verbatim}
Model: TWO-NN
Method: Least Squares Estimation
Sample size: 1500, Obs. used: 1485. Trimming proportion: 1%
ID estimates (confidence level: 0.95)

| Lower Bound| Estimate| Upper Bound|
|-----------:|--------:|-----------:|
|    1.438657| 1.456325|    1.473992|
\end{verbatim}
\begin{verbatim}
R> summary(twonn(mus = mus_gm, method = "mle"))
\end{verbatim}
\begin{verbatim}
Model: TWO-NN
Method: MLE
Sample size: 1500, Obs. used: 1485. Trimming proportion: 1%
ID estimates (confidence level: 0.95)

| Lower Bound| Estimate| Upper Bound|
|-----------:|--------:|-----------:|
|    1.739329| 1.830063|    1.925601|
\end{verbatim}

\begin{figure}[t!]
	\centering
	\includegraphics[width = .9\linewidth]{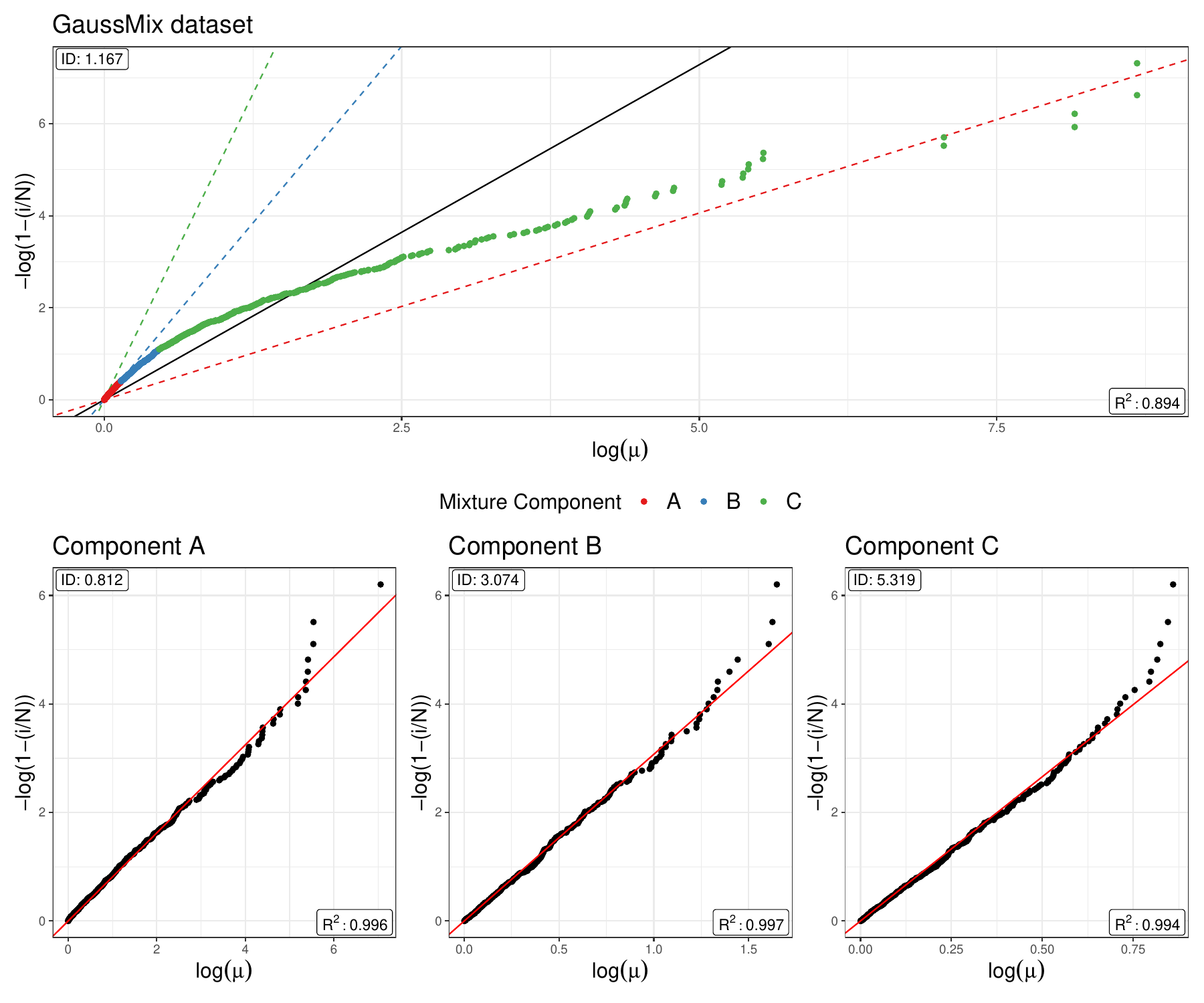}
	\caption{Linear estimators applied to the GaussMix dataset. In the top panel, the estimator is applied to the entire dataset. The points are colored according to the mixture component from which they originate, and the corresponding colored dashed lines result from applying the estimator to each known subset of points. The bottom three panels explicitly report the estimates within each mixture component.}
	\label{fig:lin_mix}
\end{figure}

The estimates obtained with the different methods do not agree.
Figure\nobreakspace{}\ref{fig:lin_mix} raises concerns about the
appropriateness of a model postulating the existence of a single, global
manifold. In the top panel of Figure\nobreakspace{}\ref{fig:lin_mix},
the data points are colored according to their generating mixture
component. The sorted log-ratios present a non-linear pattern, where the
slope values vary across the different mixture components \texttt{A},
\texttt{B}, and \texttt{C}.

\begin{figure}[t!]
	\centering
	\includegraphics[width = \linewidth]{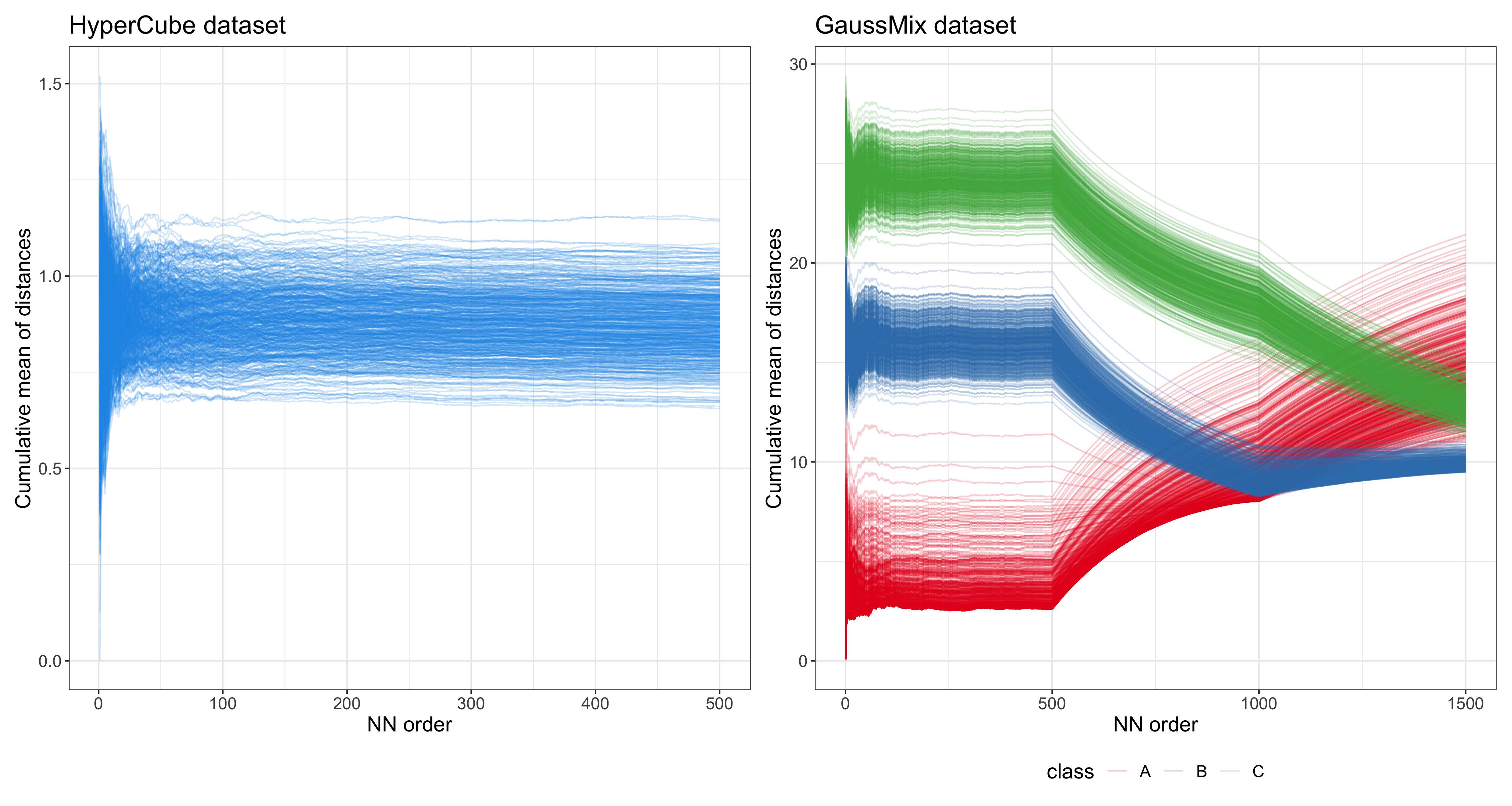} 
	\caption{Evolution of the cumulative means of NN distances computed for all the observations in the Hypercube (left panel)  and GaussMix (right panel) datasets. In the right panel, the colors highlight the different mixture components.}
	\label{fig:evolution}
\end{figure}

We can get another empirical assessment by inspecting the evolution of
the cumulative average distances between a point and its nearest
neighbors. More formally, for each point \(x_i\) we consider the
evolution of the ergodic mean of the sorted sequence of distances
\(\left(r_{i,1},\ldots,r_{i,n}\right)\), given by
\(r_i(j) = \sum_{k=1}^j r_{i,k}/k, \:\forall i,j\).
Figure\nobreakspace{}\ref{fig:evolution} compares the Hypercube (left
panel -- exemplifying the homogeneous case) and GaussMix (right panel --
representing the heterogeneous case) datasets. On the one hand, the left
plot displays the ideal condition: there are no visible departures from
the overall mean distance, which reassures us about the homogeneity
assumption we made about the data. But, on the other hand, the right
panel tells a different story. We immediately detect the different
clusters in the data by focusing on their different starting values. The
ergodic means remain approximately constant until the 500th NN, which
corresponds to the size of each subgroup. The behavior of the cumulative
means abruptly changes after the 500th NN, negating the presence of a
unique manifold. This type of plot provides an empirical but valuable
overview of the structure between data points, highlighting clusters
that may reflect the existence of multiple manifolds.

These visual assessments help detect signs of the inappropriateness of
the global ID assumption. The most direct approach to adopt in this case
would be to divide the dataset into homogeneous subgroups and apply the
TWO-NN estimator within each cluster. Such an approach is highlighted in
the bottom panels of Figure\nobreakspace{}\ref{fig:lin_mix}. However,
knowing ex-ante such well-separated groups is not a realistic
expectation to have about actual data. Therefore, we will rely on
\texttt{Hidalgo()}, the Bayesian finite mixture model for heterogeneous ID
estimation described in Section\nobreakspace{}\ref{sec:hidalgo}.

\subsubsection{Fitting the HIDALGO model}

HIDALGO allows for the presence of multiple manifolds in the same
dataset, yielding a vector of different estimated ID values. As already
discussed, estimating a mixture model with Pareto components is
challenging because of their extensive overlap. A naive model-based
estimation can lead to inaccurate results since there is no clear
separation between the kernel densities. The extra term
\(\prod_{i=1}^n\pi(\mathcal{N}_{i}^{(q)}|\bm{z})\) added into the
likelihood in Equation\nobreakspace{}\ref{MODpara} induces local
homogeneity, which helps identify the model parameters.

The adjacency matrix \(\mathcal{N}^{(q)}\) can be easily computed by
specifying two additional arguments in the function
\texttt{compute\_mus()}:

\begin{itemize}
	\item \texttt{Nq}: logical, if \texttt{TRUE}, the function adds the adjacency matrix to the output;
	\item    \texttt{q}: integer, the number of NNs to be considered in the construction of the matrix $\mathcal{N}^{(q)}$. The default value is 3.
\end{itemize}

\begin{figure}[t!]
	\centering
	\includegraphics[width=\linewidth]{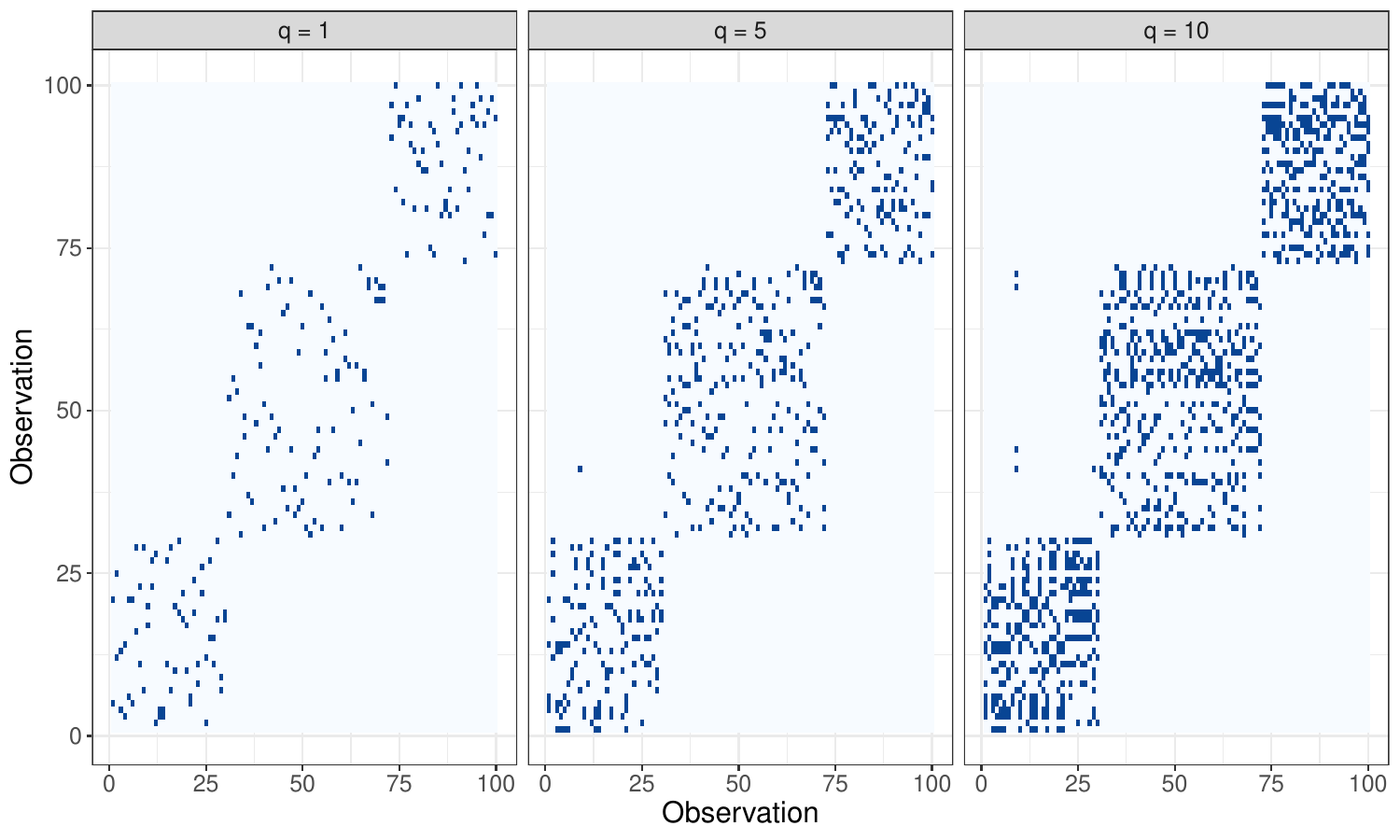} 
	\caption{Heatmaps of the adjacency matrices $\mathcal{N}^{(q)}$ computed on a subset of observations of the GaussMix dataset. Different values of $q$ are assumed.}
	\label{fig:Nqs}
\end{figure}

To provide an idea of the structure of the adjacency matrix
\(\mathcal{N}^{(q)}\), we report three examples obtained from a random
sub-sample of the GaussMix dataset for increasing values of \texttt{q}.
We display the heatmaps of the resulting matrices in
Figure\nobreakspace{}\ref{fig:Nqs}.

\begin{verbatim}
R> set.seed(12345)
R> ind <- sort(sample(1:1500, 100, F))
R> Nq1 <- compute_mus(GaussMix[ind, ], Nq = T, q = 1)$NQ
R> Nq2 <- compute_mus(GaussMix[ind, ], Nq = T, q = 5)$NQ
R> Nq3 <- compute_mus(GaussMix[ind, ], Nq = T, q = 10)$NQ
\end{verbatim}

As \texttt{q} increases, the binary matrix becomes more populated,
uncovering the neighboring structure of the data points. \cite{Allegra}
investigated how the performance of the model changes as \texttt{q}
varies. They suggest fixing \(\texttt{q}=3\), a value that provides a
good trade-off between the flexibility of the mixture allocations and
local homogeneity.

Given this premise, we are now ready to discuss \texttt{Hidalgo()}, the
high-level function that fits the Bayesian mixture. It implements the
Gibbs sampler described in Section C of the Appendix, relying on
low-level \texttt{Rcpp} routines. Also, the function internally calls
\texttt{compute\_mus()} to automatically generate the ratios of distances
and the adjacency matrix needed to evaluate the likelihood from the data
points.

The function has the following arguments: \texttt{X},
\texttt{dist\_mat}, \texttt{q}, \texttt{D}, and

\begin{itemize}
	\item   \texttt{K}: integer, number of mixture components;
	\item   \texttt{nsim}, \texttt{burn\_in}, and \texttt{thinning}: number of MCMC iterations to collect, initial iterations to discard, and thinning interval, respectively;
	\item   \texttt{verbose}: logical, if \texttt{TRUE}, the progress of the sampler is printed;
	\item   \texttt{xi}: real number between 0 and 1, a local homogeneity parameter. The default is 0.75;
	\item   \texttt{alpha\_Dirichlet}: hyperparameter of the Dirichlet prior on the mixture weights;
	\item   \texttt{a0\_d} and \texttt{b0\_d}: shape and rate parameters of the Gamma prior on $d$. The default is 1 for both values;
	\item   \texttt{prior\_type}: string, type of Gamma prior on $d$ which can be
	\begin{itemize}
\item   \texttt{"Conjugate"}: a classic Gamma prior is adopted (default);
\item   \texttt{"Truncated"}: a truncated Gamma prior on the interval $\left(0,D\right)$ is used. This specification is advised when dealing with datasets characterized by a small number of columns, to avoid the estimated ID exceeding the nominal dimension \texttt{D};
\item   \texttt{"Truncated\_PointMass"}: same as \texttt{Truncated}, but a point mass is placed on $D$. That is, the estimated ID is allowed to be exactly equal to the nominal dimension \texttt{D};
	\end{itemize}
	\item   \texttt{pi\_mass}: probability placed a priori on $D$ when a \texttt{Truncated\_PointMass} prior specification is chosen.
\end{itemize}

We apply the HIDALGO model on the GaussMix dataset with two different
prior configurations: conjugate and truncated with point mass at
\(\texttt{D}=5\). The code we used to run the models is:

\begin{verbatim}
R> set.seed(1234)
R> hid_fit <- Hidalgo(X = GaussMix, K = 10,  alpha_Dirichlet = .05,
+                    nsim = 2000,   burn_in = 2000,  thinning = 5,
+                    verbose = FALSE)
R> set.seed(12345)
R> hid_fit_TR <- Hidalgo(X = GaussMix, K = 10, alpha_Dirichlet = .05,
+                       prior_type = "Truncated_PointMass", D = 5,
+                       nsim = 2000, burn_in = 2000, thinning = 5,
+                       verbose = FALSE)
\end{verbatim}

We can print one of the returned objects to visualize a short summary of
the run:

\begin{verbatim}
R> hid_fit_TR
\end{verbatim}
\begin{verbatim}
Model: Hidalgo
Method: Bayesian Estimation
Prior d ~ Gamma(1, 1), type = Truncated_PointMass
Prior on mixture weights: Dirichlet(0.05) with 10 mixture components
MCMC details:
Total iterations: 4000, Burn in: 2000, Thinning: 5
Used iterations: 2000
Elapsed time: 2.2978 mins
\end{verbatim}

By using \texttt{alpha\_Dirichlet = 0.05}, we have adopted a sparse
mixture modeling approach in the spirit of \citet{Malsiner-Walli2016}.
The sparse mixture approach would automatically let the data estimate
the number of mixture components required. As a consequence, the
argument \texttt{K} should be interpreted as an upper bound on the number
of active clusters. Nonetheless, we stress that estimating the number of
well-separated clusters with Pareto kernels is challenging. Hence, we
will discuss how to analyze the output to perform proper inference. The
output object \texttt{hid\_fit} is a list of class \texttt{Hidalgo},
containing six elements:

\begin{itemize}
	\item  \texttt{cluster\_prob}: matrix of dimension \texttt{nsim}$\times$\texttt{K}. Each column contains the MCMC sample of a mixing weight for every mixture component;
	\item  \texttt{membership\_labels}: matrix of dimension \texttt{nsim}$\times$\texttt{n}. Each column contains the MCMC sample of a membership label for every observation;
	\item  \texttt{id\_raw}: matrix of dimension \texttt{nsim}$\times$\texttt{K}. Each column contains the MCMC sample for the ID estimated in every cluster;
	\item \texttt{id\_postpr}: matrix of dimension \texttt{nsim}$\times$\texttt{n}. It contains a chain for each observation, corrected for label-switching;
	\item \texttt{id\_summary}: a matrix containing the posterior mean and the 5\%, 25\%, 50\%, 75\%, 95\% quantiles for each observation;
	\item \texttt{recap}: a list with the specifications passed to the function as inputs.
\end{itemize}

To inspect the output, we can employ the dedicated \texttt{autoplot()}
function devised for objects of class \texttt{Hidalgo}. There are
several arguments that can be specified, producing different graphs. The
most important is

\begin{itemize}
	\item \texttt{type}: string that indicates the type of plot that is requested. It can be:
	\begin{itemize}
\item \texttt{"raw\_chains"}: plot the MCMC and the ergodic means \textbf{not} corrected for label-switching (default);
\item \texttt{"point\_estimates"}: plot the posterior mean and median ID for each observation, along with their CRIs;
\item \texttt{"class\_plot"}: plot the estimated ID distributions stratified by the groups specified in an additional \texttt{class} vector;
\item  \texttt{"clustering"}: plot the posterior co-clustering matrix. Rows and columns can be stratified by and exogenous \texttt{class} and/or a clustering structure.
	\end{itemize}
\end{itemize}

\begin{figure}[t!]
	\centering
	\includegraphics[width = .9\linewidth]{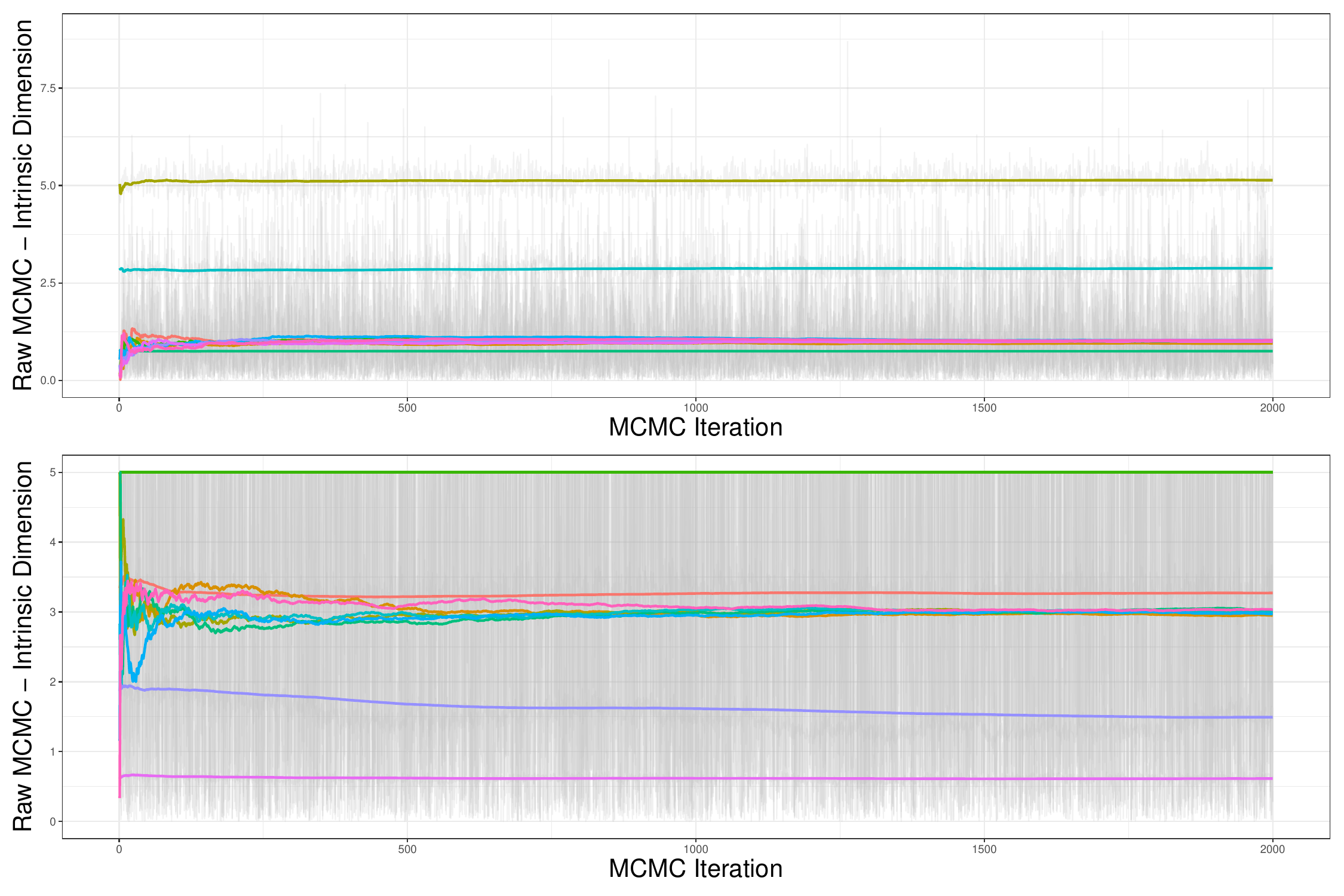}
	\caption{MCMC traceplots and superimposed ergodic means of the components of the ID vector. Top panel: conjugate prior specification. Bottom panel: truncated with point mass prior specification.}
	\label{fig:swap}
\end{figure}

For example, we can plot the raw chains of the two models with the aid
of the \texttt{patchwork} package \citep{patch}, producing
Figure\nobreakspace{}\ref{fig:swap}, via:

\begin{verbatim}
R> autoplot(hid_fit) / autoplot(hid_fit_TR)
\end{verbatim}

Plotting the traceplots of the elements in \(\bm{d}\) allows us to
assess the convergence of the algorithm. First, however, we need to be
aware that these chains may suffer from label-switching issues,
preventing us from directly drawing inference from the MCMC output. Due
to label-switching, mixture components can be discarded, emptied, or
repopulated across iterations. This behavior is observed in
Figure\nobreakspace{}\ref{fig:swap}, which shows the MCMC traceplots of
the two models, with the ergodic means for each mixture component
superimposed. In this type of plot, we can often notice that various
chains overlap around the prior mean of \(d_k\). These chains represent
the parameters of the empty clusters, which are sampled from the prior.
For example, in the top panel of Figure\nobreakspace{}\ref{fig:swap}
(\texttt{Conjugate} prior), \(\mathbb{E}\left[d_k\right]=a_d/b_d=1\).
Recall that the presence of empty clusters is favored by the sparse
mixture setting.\\
Additionally, we can see that if no constraint is imposed on the support
of the prior distribution for \(\bm{d}\) (top panel), the posterior
estimates can exceed the nominal dimension \texttt{D = 5} of the
GaussMix dataset. However, this problem disappears when imposing a
truncation on the prior support (bottom panel).

To address the label-switching issue and perform meaningful inference,
the raw MCMC needs to be postprocessed. In Section D of the Appendix, we
discuss the algorithm used to map the \(K\) chains to \(n\)
observation-specific chains that can be employed for inference. The
algorithm is already implemented in \texttt{Hidalgo()}, and produces the
elements \texttt{id\_postpr} and \texttt{id\_summary} in the returned list.
We can obtain a visual summary of the postprocessed estimates via

\begin{verbatim}
R> autoplot(hid_fit, type = "point_estimates") +
+  autoplot(hid_fit_TR, type = "point_estimates")
\end{verbatim}

\begin{figure}[t!]
	\centering
	\includegraphics[width = .95\linewidth]{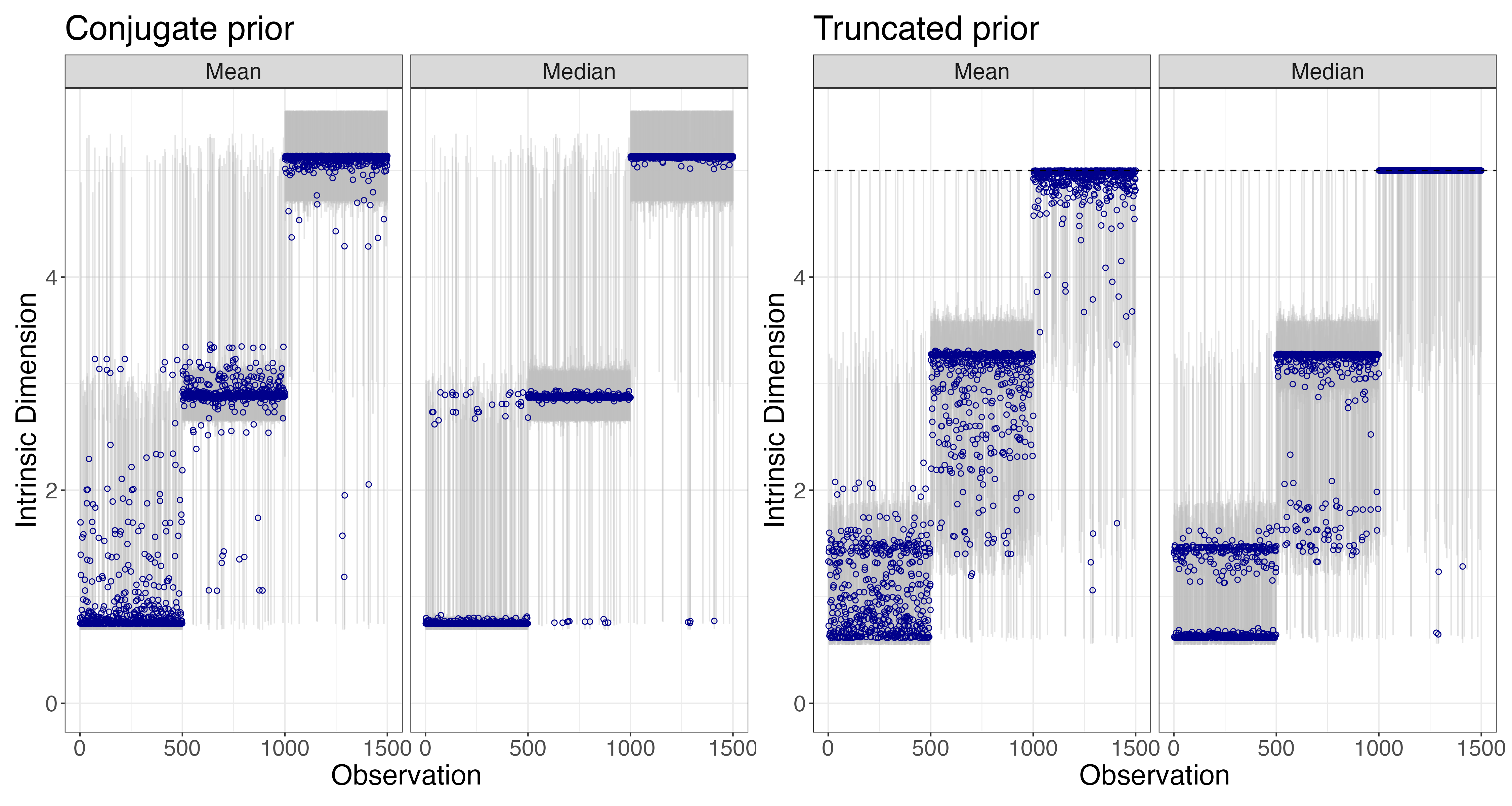}
	\caption{Observation-specific posterior means (left panels) and medians (right panels) ID represented with blue dots. The gray bars represent the $90\%$ CRIs. The two plots correspond to two different prior specifications.}
	\label{fig:memed}
\end{figure}

The resulting plots are shown in Figure\nobreakspace{}\ref{fig:memed}.
The panels display the mean and median ID estimates for each data point.
Here, the separation of the data into different generating manifolds is
evident. Also, we notice that some of the estimates in the conjugate
case are incorrectly above the nominal value \texttt{D = 5}, once again
justifying the need for a truncated prior. The default plots were
modified with \texttt{coord\_cartesian(ylim = c(0, 5.5))} to highlight
the effect of the truncation.

\subsubsection{Estimated clustering solutions}

It is natural to seek model-based clustering solutions when dealing with
a mixture model. To this extent, the key source of information is the
posterior similarity -- or co-clustering -- matrix (PSM). The entries
\(\{s_{i,j}\}_{i,j=1}^n\) of this matrix are computed as the proportion
of times in which two observations have been assigned to the same
mixture component across the MCMC iterations. Thus, the PSM describes
the underlying clustering structure of the data detected by HIDALGO.
Given the PSM, one can evaluate various loss functions on the space of
the partitions. By minimizing the loss functions, we can retrieve the
optimal partition of the dataset into clusters. To obtain such estimate,
we rely on the function \texttt{salso()} from the \texttt{R} packages
\texttt{salso} \citep{salso_package}. Otherwise, a faster alternative
method proceeds by building a dendrogram from the implied posterior
dissimilarity matrix (PDM), whose entries are given by
\(\{d_{i,j}\}_{i,j=1}^n\) where \(d_{i,j}=1-s_{i,j}\), \(\forall i,j\).
Once the dendrogram is built, we can threshold it to segment the data
into a pre-specified number of clusters \texttt{K}.

These approaches are implemented in the dedicated function
\texttt{clustering()} which takes as arguments, along the \texttt{object}
output from the \texttt{Hidalgo()} function,

\begin{itemize}
	\item \texttt{clustering\_method}: string indicating the method to use to perform clustering. It can be \texttt{"dendrogram"} or \texttt{"salso"}. The former method thresholds the dendrogram constructed from the PDM to retrieve exactly \texttt{K} clusters. The latter method estimates the optimal clustering solution by minimizing a loss function on the space of the partitions. The default loss function is the variation of information \citep[VI,][]{Wade2018}. For additional details about the VI loss function, see Section E of the Appendix;
	\item \texttt{K}: integer, used when \texttt{"dendrogram"} is chosen. It corresponds to the number of clusters to recover when thresholding the dendrogram obtained from the PDM;
	\item \texttt{nCores}: integer, argument used in the functions called from \texttt{salso}. It represents the number of cores used to compute the PSM and the optimal clustering solution. 
\end{itemize}

Additional arguments can be passed to personalize the partition
estimation via \texttt{salso()}. Given the large sample size of the
GaussMix dataset, we opt for the dendrogram approach, truncating the
dendrogram at \texttt{K = 3} groups. We highlight that relying on the
minimization of a loss function is a more principled approach. However,
the method can be misled by the strongly overlapping clusters estimated
across the MCMC iterations, providing overly conservative solutions.

\begin{verbatim}
R> psm_cl <- clustering(object = hid_fit_TR, 
+                      clustering_method = "dendrogram", 
+                      K=3, nCores = 5)
R> psm_cl
\end{verbatim}
\begin{verbatim}
Estimated clustering solution summary:

Method: dendrogram
Retrieved clusters: 3
Clustering frequencies:

| Cluster 1| Cluster 2| Cluster 3|
|---------:|---------:|---------:|
|       554|       450|       496|
\end{verbatim}

To visualize the results, we can also plot the PSM by passing an object
of class \texttt{Hidalgo} to \texttt{autoplot()} with
\texttt{type = "clustering"}. \texttt{autoplot()} internally calls the
function \texttt{clustering()} to compute the PSM. One can also specify an
additional argument \texttt{class} to stratify the observations
according to exogenous factors.

\subsubsection{The presence of patterns in the data uncovered by the ID}

Once the observation-specific estimates are computed, we can investigate
the presence of potential patterns between the recovered IDs and given
exogenous variables. To explore these possible relations, we can use the
function \texttt{id\_by\_class()}. Along with an object of class
\texttt{Hidalgo}, we need to specify:

\begin{itemize}
	\item  \texttt{class}: factor, a variable used to stratify the ID posterior estimates.
\end{itemize}

For the GaussMix dataset, the exogenous information is contained in the
\texttt{class\_GMix} vector, which we pass as \texttt{class}.

\begin{verbatim}
R> id_by_class(object = hid_fit_TR, class = class_GMix)
\end{verbatim}
\begin{verbatim}
Posterior ID by class:

|class |     mean|    median|        sd|
|:-----|--------:|---------:|---------:|
|A     | 1.031105| 0.9019793| 0.3781802|
|B     | 2.961009| 3.2068593| 0.4683262|
|C     | 4.864018| 4.9685900| 0.3774186|
\end{verbatim}

The estimates in the three classes are very close to the ground truth.
The same argument, \texttt{class}, can be passed to the
\texttt{autoplot()} function, in combination with

\begin{figure}[t!]
	\centering
	\includegraphics[width = .9\linewidth]{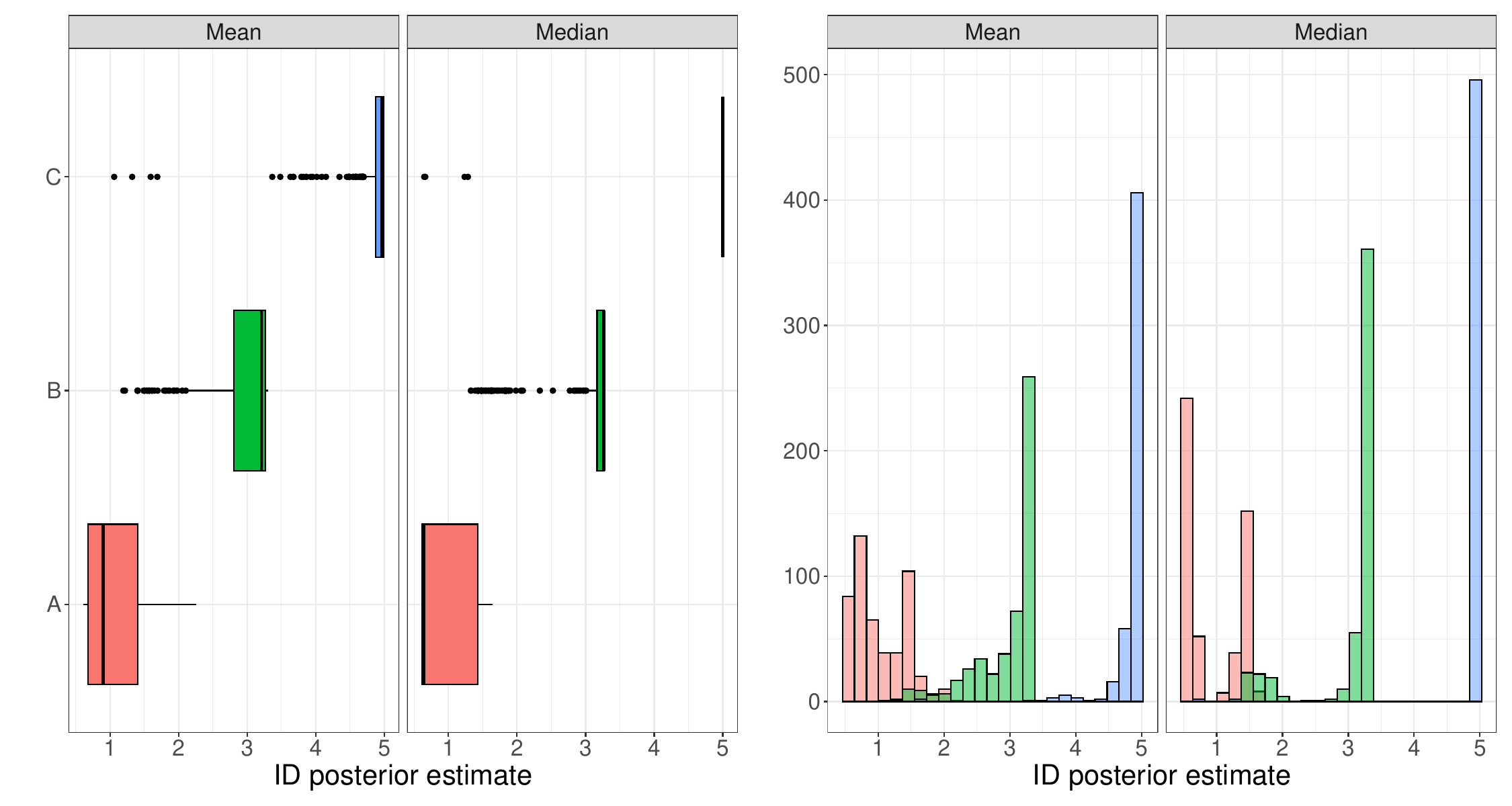} 
	\caption{Two different types of graphs used to stratify the estimated IDs by a given exogenous variable.}
	\label{fig:classes}
\end{figure}

\begin{itemize}
	\item \texttt{class\_plot\_type}: string, if \texttt{type = "class\_plot"}, one can visualize the stratified ID estimates with a \texttt{"density"} plot or a \texttt{"histogram"}, or using \texttt{"boxplots"} or \texttt{"violin"} plots;
	\item  \texttt{class}: a vector containing a class used to stratify the observations;
\end{itemize}

to visualize ID estimates of the GaussMix dataset stratified by the
generating manifold of the observations. As an example of possible
graphs, Figure\nobreakspace{}\ref{fig:classes} shows the stratified
boxplots (left panel) and histograms (right panel).

\begin{verbatim}
R> autoplot(hid_fit_TR, type = "class", class = class_GMix, 
+          class_plot_type = "boxplot") +
+ autoplot(hid_fit_TR, type = "class", class = class_GMix, 
+           class_plot_type = "histogram")
\end{verbatim}

We have introduced and discussed the principal functions of the
\texttt{intRinsic} package concerning the TWO-NN and HIDALGO models. Then,
by employing simulated data with known IDs, we have suggested a pipeline
to guide our study. In the next section, we present a real data
analysis, highlighting how the ID estimation can be used to effectively
reduce the size of a dataset while capturing and preserving important
features.

\section{The ID of gene microarray measurements}
\label{sec:alon}

\begin{figure}[ht!]
	\centering
	\includegraphics[width=\linewidth]{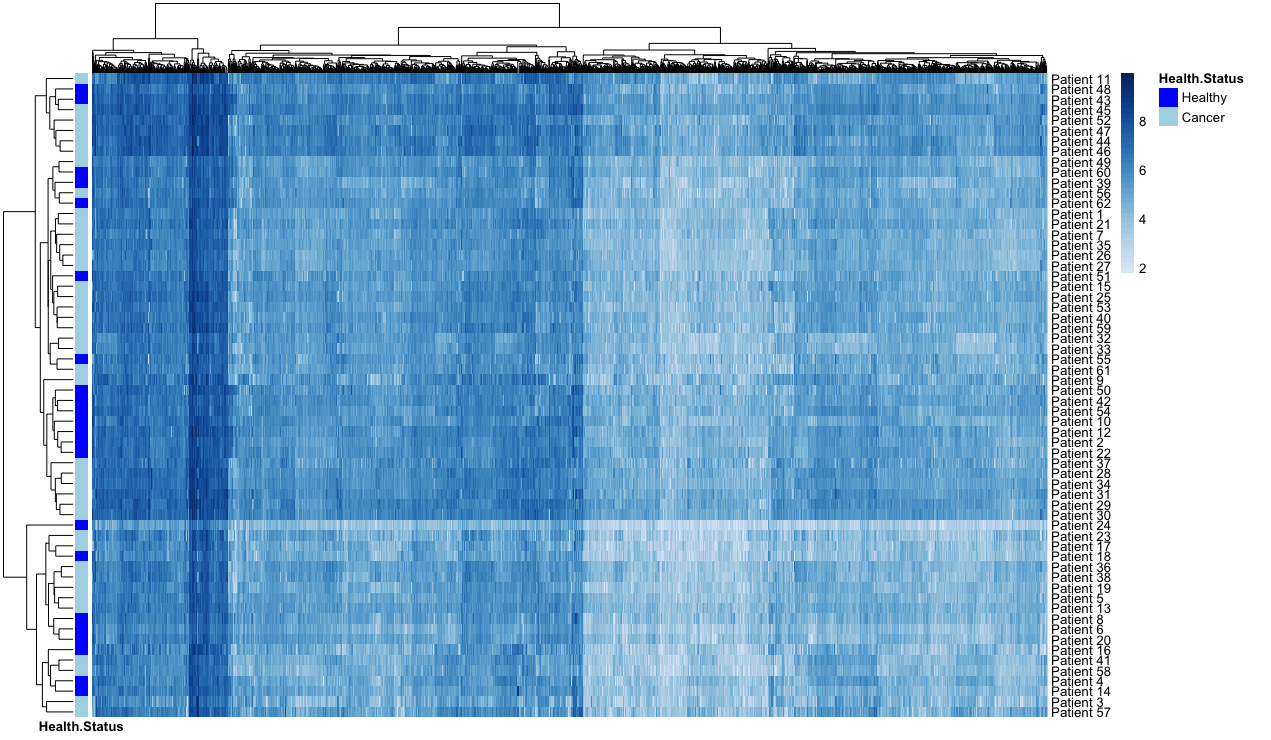}
	\caption{Heatmap of the log-values of the Alon microarray dataset. The patients on the rows are labeled according to their health status.}
	\label{fig:Alon}
\end{figure}

In this section, we present a real data example investigating the ID of
the Alon dataset. The dataset, first presented in \citet{Alon6745},
contains microarray measurements for 2000 genes measured on 62 patients.
Among the patients, 40 were diagnosed with colon cancer, and 22 were
healthy subjects. A factor variable named \texttt{status} describes the
patient health condition (coded as \texttt{"Cancer"}
vs.~\texttt{"Healthy"}). A copy of this famous dataset can be found in
the \texttt{R} package \texttt{HiDimDA} \citep{RHiDimDA}. We store the
gene measurements in the object \texttt{Xalon}, a matrix of nominal
dimension \texttt{D = 2000}, with \texttt{n = 62} observations. To load
and prepare the data, we write:

\begin{verbatim}
R> data("AlonDS", package = "HiDimDA")
R> status <- factor(AlonDS$grouping, labels = c("Cancer", "Healthy"))
R> Xalon  <- as.matrix(AlonDS[, -1])
\end{verbatim}

To obtain a visual summary of the dataset, we plot the heatmap of the
log-data values annotated by \texttt{status}. The result is shown in
Figure\nobreakspace{}\ref{fig:Alon}. No clear structure is immediately
visible.

We ultimately seek to uncover hidden patterns in this dataset. The task
is challenging, especially given the small number of available
observations. As a first step, we investigate how well a unique, global
ID estimate can represent the data.

\subsection{Homogeneous ID estimation}

Let us start by describing the overall complexity of the dataset by
estimating a homogeneous ID value. Using the TWO-NN model, we can
compute:

\begin{verbatim}
R> Alon_twonn_1 <- twonn(Xalon,method = "linfit")
R> summary(Alon_twonn_1)
\end{verbatim}
\begin{verbatim}
Model: TWO-NN
Method: Least Squares Estimation
Sample size: 62, Obs. used: 61. Trimming proportion: 1%
ID estimates (confidence level: 0.95)

| Lower Bound| Estimate| Upper Bound|
|-----------:|--------:|-----------:|
|    10.00382| 10.34944|    10.69506|
\end{verbatim}
\begin{verbatim}
R> Alon_twonn_2 <- twonn(Xalon,method = "bayes")
R> summary(Alon_twonn_2)
\end{verbatim}
\begin{verbatim}
Model: TWO-NN
Method: Bayesian Estimation
Sample size: 62, Obs. used: 61. Trimming proportion: 1%
Prior d ~ Gamma(0.001, 0.001)
Credibile Interval quantiles: 2.5%, 97.5%
Posterior ID estimates:

| Lower Bound|     Mean|   Median|     Mode| Upper Bound|
|-----------:|--------:|--------:|--------:|-----------:|
|    7.784152| 10.17639| 10.12084| 10.00957|    12.88427|
\end{verbatim}
\begin{verbatim}
R> Alon_twonn_3 <- twonn(Xalon,method = "mle")
R> summary(Alon_twonn_3)
\end{verbatim}
\begin{verbatim}
Model: TWO-NN
Method: MLE
Sample size: 62, Obs. used: 61. Trimming proportion: 1%
ID estimates (confidence level: 0.95)

| Lower Bound| Estimate| Upper Bound|
|-----------:|--------:|-----------:|
|    7.785304| 10.01107|    12.88623|
\end{verbatim}

\begin{figure}[t!]
	\centering
	\includegraphics[width=\linewidth]{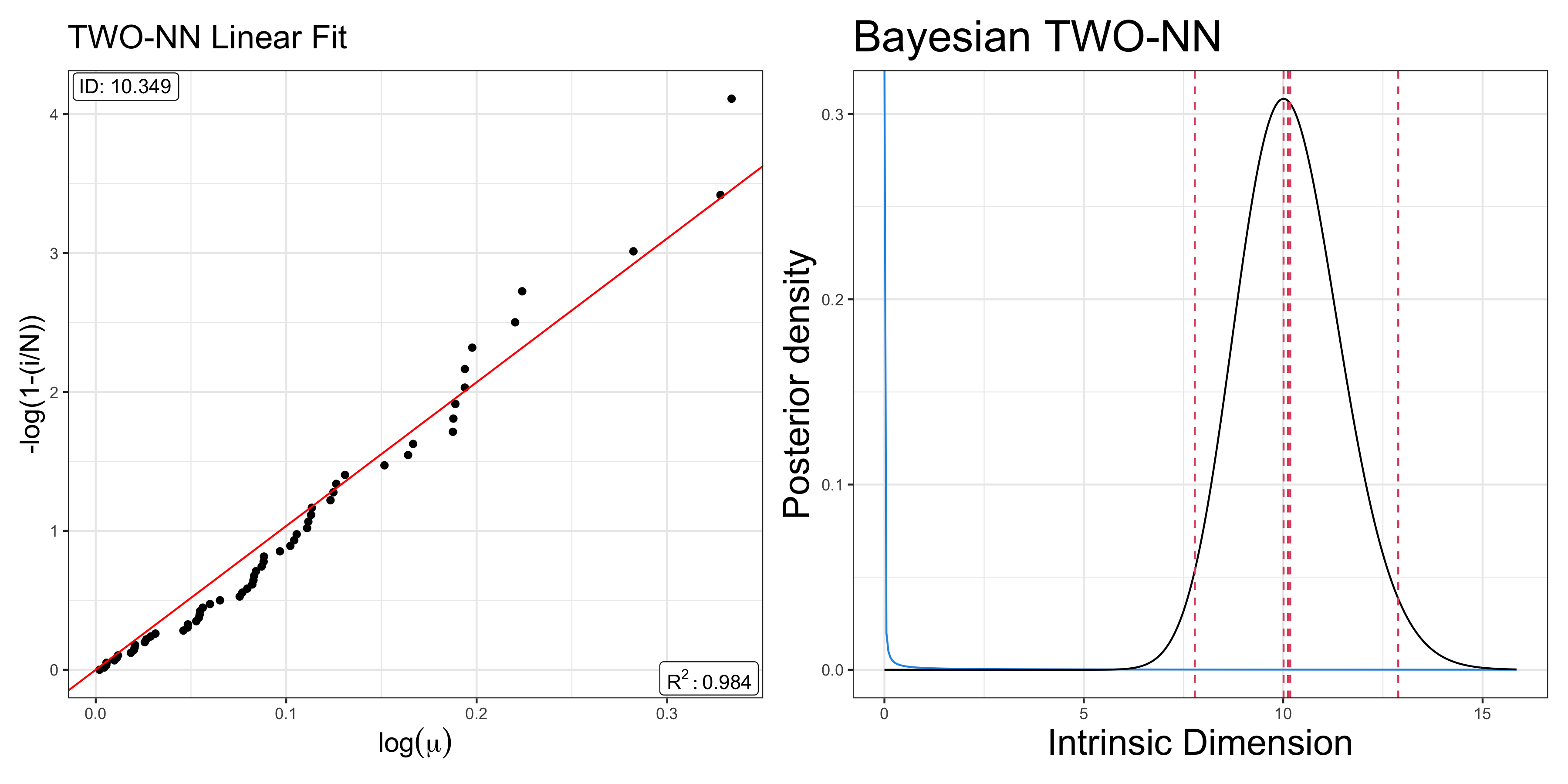}
	\caption{Alon dataset. The left panel shows the result of the linear estimator, while the right panel depicts the posterior distribution obtained via the Bayesian approach.}
	\label{fig:linbay}
\end{figure}

The estimates based on the TWO-NN model obtained with different methods
are very similar. The results are also illustrated in
Figure\nobreakspace{}\ref{fig:linbay}, which shows the linear fit (left
panel) and posterior distribution (right panel) for the TWO-NN model.
According to these results, we conclude that the information contained
in the \texttt{D = 2000} genes can be summarized with approximately ten
variables. For example, the first ten eigenvalues computed from the
spectral decomposition of the matrix
\(\Lambda = X_{Alon}^{'}X^{\:}_{Alon}\) contribute to explaining the
95.4\% of the total variance.

\begin{figure}[th!]
	\centering
	\includegraphics[width=\linewidth]{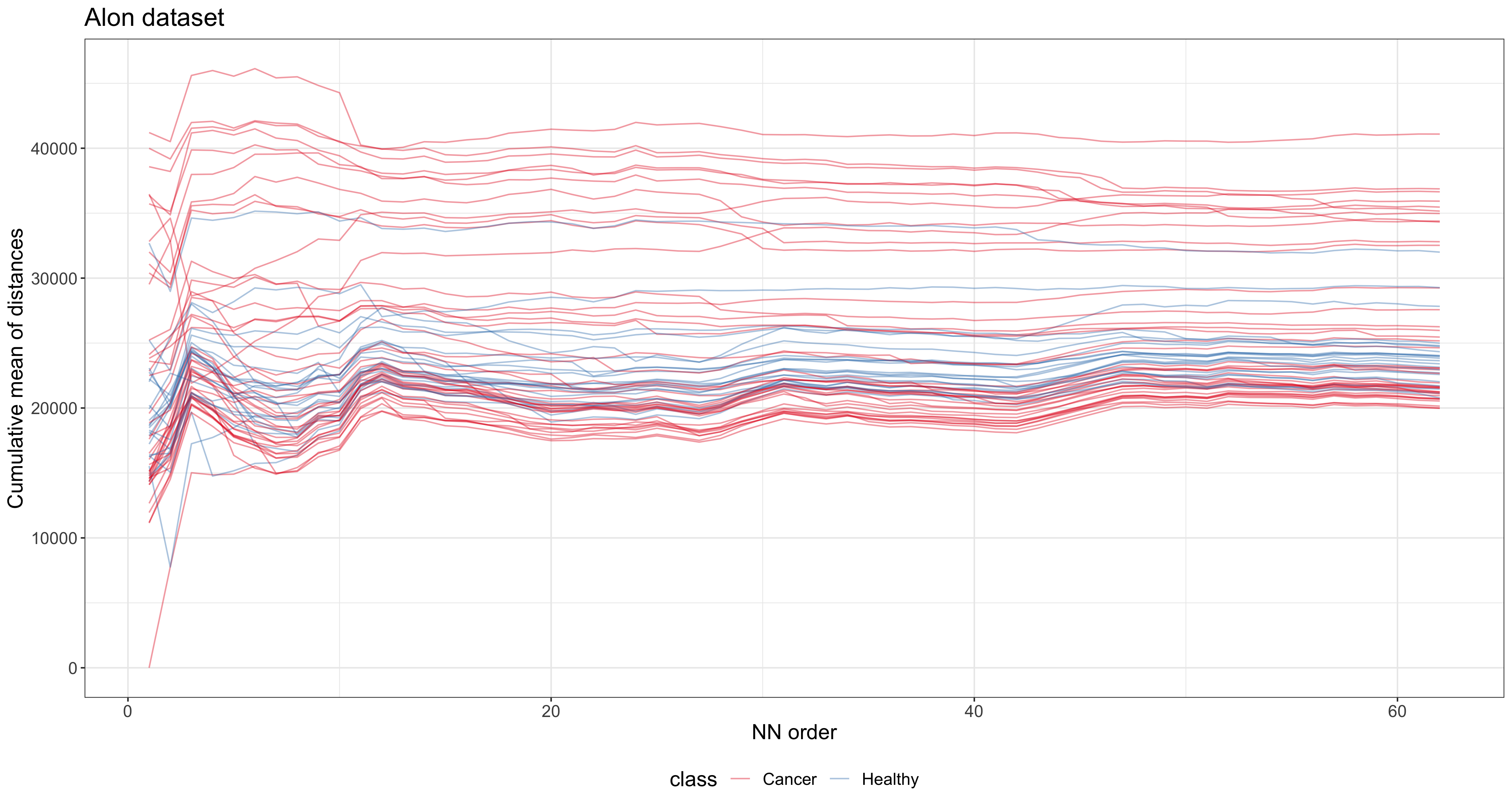}
	\caption{Evolution of the cumulative means of NN distances computed for all the observations in the Alon dataset.}
	\label{fig:evo_alon}
\end{figure}

Although the linear fit plot and the TWO-NN estimates do not raise any
evident sign of concern, as a final check, we explore the evolution of
the average distances between NN, reported in
Figure\nobreakspace{}\ref{fig:evo_alon}. As expected, the plot does not
highlight any abrupt change in the evolution of the ergodic means.
However, it suggests that investigating the presence of multiple
manifolds could be interesting. In fact, despite the evolution of most
of the ergodic means being stationary, their heterogeneous levels
highlight some potential data inhomogeneities that should be deepened.

\subsection{Heterogeneous ID estimation}

To investigate the presence of heterogeneous latent manifolds in the
Alon dataset, we employ \texttt{Hidalgo()}. Since the nominal dimension
\texttt{D} is large, we do not need to truncate the prior on \(d\).
Moreover, given the small number of data points, we opt for an
informative and regularizing prior \(Gamma(1,1)\) (the default) instead
of a vague specification. Also, we set a conservative upper bound for
the mixing component \texttt{K = 15} and choose again \(\alpha = 0.05\)
to fit a sparse mixture. We run:

\begin{verbatim}
R> set.seed(1234)
R> Alon_hid <- Hidalgo(X = Xalon,  K = 15, a0_d = 1, b0_d = 1, 
+             alpha_Dirichlet = .05, 
+             nsim = 10000, burn_in = 100000, thin = 5)
\end{verbatim}

\begin{verbatim}
R> Alon_hid
\end{verbatim}
\begin{verbatim}
Model: Hidalgo
Method: Bayesian Estimation
Prior d ~ Gamma(1, 1), type = Conjugate
Prior on mixture weights: Dirichlet(0.05) with 15 mixture components
MCMC details:
Total iterations: 110000, Burn in: 1e+05, Thinning: 5
Used iterations: 10000
Elapsed time: 38.3067 secs
\end{verbatim}

\begin{figure}[t]
	\centering
	\includegraphics[width = \linewidth]{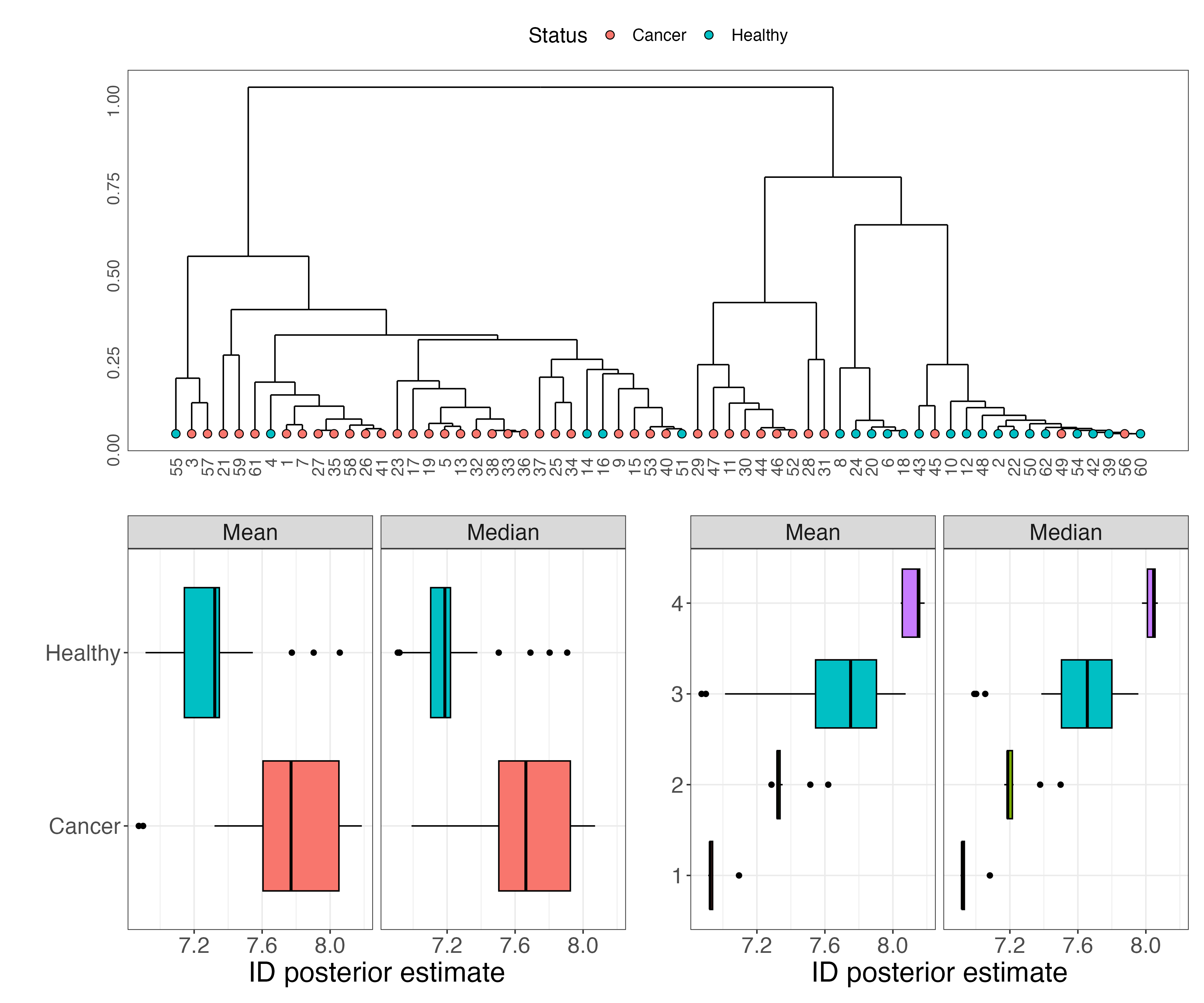}
	\caption{Alon dataset. Top panel: dendrogram obtained from the PDM. Bottom panel: boxplots of the ID estimates stratified by health status (left) and estimated partition (right).}
	\label{fig:dendro}
\end{figure}

Once the model is fitted, we first explore the estimated clustering
structure. Here, instead of directly plotting the heatmap of the PSM, we
build the dendrogram from the PDM, and we report it in the top panel of
Figure\nobreakspace{}\ref{fig:dendro}. We construct such a plot with the
help of the package \texttt{ggdendro} \citep{ggdendro}. We can detect four
clusters, and therefore we decide to set \texttt{K = 4} when running

\begin{verbatim}
R> Alon_psm <- clustering(Alon_hid, K = 4)
\end{verbatim}

As illustrated in the previous section, all these plots can be obtained
by calling \texttt{autoplot()} with the proper argument specifications.
The next natural step is to investigate how strongly the estimated
partition and, in general, the estimated IDs are associated with health
\texttt{status}. The bottom two panels of
Figure\nobreakspace{}\ref{fig:dendro} display the boxplots of the values
(means and medians) of the postprocessed, observation-specific IDs
stratified by \texttt{status} (left) and estimated cluster (right). We
can also run

\begin{verbatim}
R> id_by_class(Alon_hid,class = status)
R> id_by_class(Alon_hid,class = Alon_psm$clust)
\end{verbatim}

to obtain summary results linking the variations in the ID with health
status and cluster. We report the output in
Table\nobreakspace{}\ref{tab:alon_summary}. As the estimated ID
increases, the proportion of healthy subjects in each cluster decreases.
This result suggests that the microarray profiles of people diagnosed
with cancer are slightly more complex than healthy patients' ones.

\begin{table}[t!]
	\centering
	\begin{tabular}{ccccccc}
\toprule
Cluster & \# \texttt{Cancer} & \# \texttt{Healthy} & \% \texttt{Healthy} & Average ID & Median ID & Std. Dev.\\
\midrule
1 & 0 & 5 & 1.0000 & 6.9595 & 6.9252 & 0.0761 \\
2 & 3 & 12 & 0.8000 & 7.3564 & 7.3222 & 0.0889\\
3 & 28 & 5 & 0.1515 & 7.6883 & 7.7513 & 0.3083\\
4 & 9 & 0 & 0.0000 & 8.1198 & 8.1493 & 0.0546 \\
\bottomrule
	\end{tabular}
	\caption{Stratification by cluster of the health status (frequencies and proportions of healthy patients) and ID estimates (mean, median, and standard deviation).}
	\label{tab:alon_summary}
\end{table}

The analyses conducted so far helped us uncover interesting descriptive
characteristics of the IDs in the dataset. Nevertheless, these results
can also be effectively used as a representative data summary. Here, we
show how the estimated individual IDs are valid to potentially classify
the health \texttt{status} of new patients according to their genomic
profiles. As a simple example, we perform a classification analysis
using two \texttt{random forest} models, predicting the target variable
\texttt{Y = status}. To train the models, we use two different sets of
covariates: \texttt{X\_OR}, the original dataset composed of 2000 genes,

\begin{verbatim}
R> X_OR <- data.frame(Y = status, X = Xalon)
R> set.seed(1231)
R> rfm1 <- randomForest::randomForest(Y ~ ., data = X_OR, 
+                                    type = "classification", ntree=100)
\end{verbatim}

and \texttt{X\_ID}, the observation-specific ID summary returned by
\texttt{Hidalgo()}, along with our estimated partition.

\begin{verbatim}
R> X_ID <- data.frame(Y     = status,
+                    X     = summary(Alon_hid),
+                    clust = factor(Alon_psm$clust))
R> set.seed(1231)
R> rfm2 <- randomForest::randomForest(Y ~ ., data = X_ID, 
+                                    type = "classification", ntree=100)
\end{verbatim}

The classification results are reported in
Table\nobreakspace{}\ref{tab:rfs}.

\begin{table}[t!]
	\centering
	\begin{tabular}{cccccccc}
\toprule
\texttt{rfm1} & \texttt{Cancer} & \texttt{Healthy} & \texttt{class.err} &  \texttt{rfm2} & \texttt{Cancer}  & \texttt{Healthy} & \texttt{class.err} \\
\midrule
\texttt{Cancer}& 36 & 4 & 0.100 &   \texttt{Cancer}& 35 & 5 & 0.1250\\
\texttt{Healthy} & 9 & 13 & 0.409 & \texttt{Healthy} & 6 & 16 & 0.273\\

\midrule
\texttt{Dataset:} & \texttt{X\_OR} & \texttt{OOB err:}& 20.97\% &
\texttt{Dataset:} & \texttt{X\_ID} & \texttt{OOB err:}& 17.74\%\\
\bottomrule
	\end{tabular}
	\caption{Confusion matrices summarizing the classification performance of the two random forest models, trained with the orginal gene expressions (\texttt{rfm1}, left) and with a summary of the ID estimates (\texttt{rfm2}, right).}
	\label{tab:rfs}
\end{table}

Remarkably, a simple dataset with seven variables summarizing the main
distributional traits of the observation-specific posterior IDs obtains
good performance in predicting health status, similar to the original
dataset. More precisely, the random forest on the original dataset got
an out-of-bag estimated error rate of 20.97\%, while the error is
reduced to 17.74\% when using our ID-based covariates. We can conclude
that, in this case, the topological properties of the dataset are
associated with the outcome of interest and convey important
information.

We showed how the estimation of heterogeneous ID provides a reliable
complexity index for elaborate data structures and helps unveil
relationships among data points hidden at the topological level. The
application to the Alon dataset showcases how reliable ID estimates give
additional fundamental perspectives that help us discover non-trivial
data patterns. Furthermore, one can exploit the extracted information in
many downstream investigations, such as patient segmentation or
predictive analyses.

\section{Summary and discussion} \label{sec:summary}

In this paper, we illustrated \texttt{intRinsic}, an \texttt{R} package
that implements novel routines for the ID estimation according to the
models recently developed in \citet{Facco, Allegra, Denti2021}, and
\citet{Santos-Fernandez2020}. \texttt{intRinsic} consists of a collection
of high-level, user-friendly functions that, in turn, rely on efficient,
low-level routines implemented in \texttt{R} and \texttt{C++}. We
also remark that \texttt{intRinsic} integrates functionalities from
external packages. For example, all the graphical outputs returned by
the functions are built using the well-known package \texttt{ggplot2}.
Therefore, they are easily customizable using the grammar of graphics
\citep{Wilkinson2005}.

The package includes frequentist and Bayesian model specifications for
the TWO-NN global ID estimator. Moreover, it implements the Gibbs
sampler for posterior simulation of the HIDALGO model, which can capture
the presence of heterogeneous ID within a single dataset. We showed how
discovering multiple latent manifolds could help unveil the topological
traits of a dataset, primarily when additional exogenous variables are
used to stratify the ID estimates.

As a general analysis pipeline for practitioners, we suggested starting
with the efficient TWO-NN functions to understand how appropriate the
hypothesis of homogeneity is for the data at hand. If departures from
the assumptions are visible from nonuniform estimates obtained with
different estimation methods and from visual assessment of the evolution
of the average NN distances, one should rely on HIDALGO.

The most promising future research directions stem from HIDALGO. First,
we plan to develop more reliable methods to obtain an optimal partition
of the data based on the ID estimates since the one proposed heavily
relies on a mixture model of overlapping distribution. Moreover, another
research avenue worth exploring is a version of HIDALGO with likelihood
distributions based on generalized NN ratios, exploiting the information
coming from varying neighborhood sizes.\\
We also know that the mixture model fitting may become computationally
expensive if the analyzed datasets are large. Therefore, faster
solutions, such as the Variational Bayes approach, will be explored.
Also, we highlight that HIDALGO, a mixture model within a Bayesian
framework, lacks a frequentist-based estimation counterpart, such as an
Expectation Maximization algorithm. Its derivation is not immediate
since the neighboring structure introduced via the \(\mathcal{N}^{(q)}\)
matrix makes the problem non-trivial. We plan to keep working on this
package and continuously update it in the long run as contributions to
this line of research become available. The novel ID estimators we
discussed have started a lively research branch, and we intend to
include all the future advancements in \texttt{intRinsic}.

\section*{Acknowledgements}

The author thanks the Editorial Team and the two anonymous Reviewers for
their constructive comments. Moreover, the author is extremely grateful
to Andrea Gilardi for his valuable guidance. Finally, the author also
thanks Michelle N. Ngo, Derenik Haghverdian, Wendy N. Rummerfield,
Andrea Cappozzo, and Riccardo Corradin for their comments on earlier
versions of this manuscript.

\section*{Appendix}

\subsection*{A - Additional methods implemented in the package}

In this paper, we have focused our attention on the TWO-NN and the
HIDALGO models. In Section\nobreakspace{}\ref{sec:mod}, we explained
that both methods are based on the distributional properties of the
ratios of distances between a point and its first two NNs. However, this
modeling framework has been extended by \citet{Denti2021}, where the
authors developed a novel ID estimator called GRIDE. This new estimator
is based upon the ratios of distances between a point two of its NNs of
generic order, namely \(n_1\) and \(n_2\). Extending the neighborhood
size leads to two major implications: more stringent local homogeneity
assumptions and the possibility of computing ID estimates as a function
of the chosen NN orders. Monitoring the ID evolution as the order of the
furthest NN \(n_2\) increases allows the extraction of meaningful
information regarding the link between the ID and the scale of the
considered neighborhood. In doing so, GRIDE produces estimates that are
more robust to noise present in the data, which is not directly
addressed by the model formulation.

The GRIDE model is implemented in \texttt{intRinsic}, and the estimation
can be carried out under both the frequentist and Bayesian frameworks
via the function \texttt{gride()}, which is very similar to \texttt{twonn()}
in its usage. Additionally, one can use the functions
\texttt{twonn\_decimation()} and \texttt{gride\_evolution()} to study the ID
dynamics. More details about these functions are available in the
package documentation.

\begin{figure}[t!]
	\centering
	\includegraphics[width = \linewidth, trim={0cm 3cm 0cm 3cm},clip]{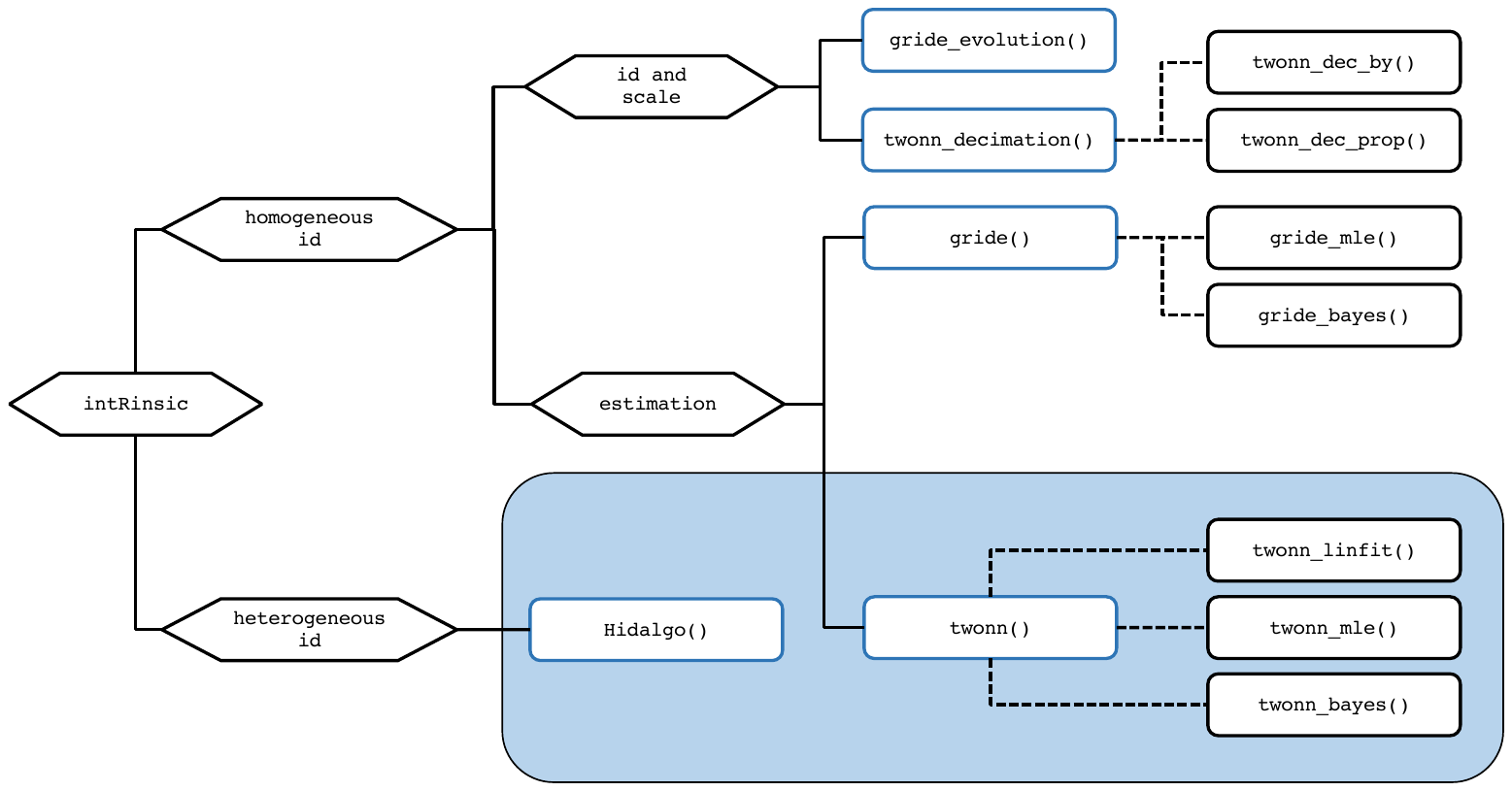}
	\caption{A conceptual map summarizing the most important functions contained in \texttt{intRinsic}. The blue squares contain the names of the principal, high-level functions. Dotted lines connect these functions with the most important low-level functions (not exported). The light-blue area represents the topics that have been discussed in this paper.}
	\label{fig:map}
\end{figure}

The map in Figure\nobreakspace{}\ref{fig:map} provides a visual summary
of the most important functions contained in the package. The main
topics are reported in the diamonds, while the high-level, exported
functions are displayed in the blue rectangles. These routines are
linked to the (principal) low-level function via dotted lines. Finally,
the light-blue area highlights the functions discussed in this paper.

\subsection*{B - \texttt{intRinsic} and other packages}

As mentioned in Section\nobreakspace{}\ref{sec:intro}, there is a large
number of ID estimators available in the literature, and many of them
have been implemented in \texttt{R}. A valuable survey of the
availability of dimensionality reduction methods and ID estimators has
been recently reported in \citet{You2020}. From there, we see that two
packages are the most important when it comes to ID estimation:
\texttt{Rdimtools} and \texttt{intrinsicDimension}. \texttt{Rdimtools}, in
particular, is comprised of an unprecedented collection of methods --
including also the least-squares TWO-NN.\\
At the moment of writing, two main traits of \texttt{intRinsic} are unique
to this package.\\
First, our package is devoted to the recently proposed likelihood-based
estimation methods introduced by the seminal work of \citet{Facco} and
the literature that followed. Therefore, as of today, many of the
\texttt{R} implementations presented here are exclusively contained in
this package. This is true, for example, for the MLE and Bayesian
versions of the TWO-NN model, the HIDALGO model, and all the routines
linked to GRIDE. To the best of our knowledge, the function
\texttt{Hidalgo()} is available outside this package. However, one can
only find it on \texttt{GitHub} repositories, coded in \texttt{Python}
and \texttt{C++}. Note that \texttt{Python} versions of the TWO-NN
estimator have also been implemented in the recent
\texttt{scikit-dimension} and \texttt{DADApy} packages \citep{Bac2021,dadapy}.
Moreover, \texttt{DADApy} contains routines dedicated to GRIDE. Table
\ref{tab:summa} presents a summary of recent software packages
containing ensembles of ID estimation methods.\\
Second, all the functions in our package allow -- and emphasize -- the
uncertainty quantification around the ID estimates, which is a crucial
component granted by our model-based approach. This feature is often
overlooked in other implementations.

\begin{table}[th!]
	\begin{tabular}{ccccc}
\toprule
Package  &Language & ID estimation methods & Overlap with \texttt{intrRinsic}\\
\midrule
\texttt{intrinsicDimension} & \texttt{R}& 5 &  --\\
\texttt{Rdimtools}  & \texttt{R}& 17 & TWO-NN$^\mathsection$\\
\texttt{ider}  & \texttt{R}& 7 & -- \\
\texttt{DADApy}  & \texttt{Python} & 2 & TWO-NN$^{\dagger,\ddagger,\mathsection}$, GRIDE$^\dagger$\\
\texttt{scikit-dimension}  & \texttt{Python} & 19 & TWO-NN$^\mathsection$\\
\bottomrule 
	\end{tabular}
	\caption{A non-exhaustive list of the most recent software packages for ID estimation in \texttt{R} and \texttt{Python}. The superscripts indicate the implemented estimation procedures. In detail: $^\dagger$MLE; $^\ddagger$Bayesian estimation; $^\mathsection$linear fit.}
	\label{tab:summa}
\end{table}

Overall, the wide variety of methods and ongoing research in this area
indicate that there is no globally optimal estimator to employ
regardless of the application. Thus, a practitioner should be aware of
the strengths and limitations of every method.\\
One limitation of the likelihood-based models offered in this package,
shared with many other ID estimators in general, is the underestimation
of the ID when the true latent manifold's dimension is large. As an
empirical rule, for cases where the estimated ID is large (e.g.,
\(d>20\)), the retrieved value should be cautiously regarded as a lower
bound for the actual ID \citep{Ansuini2019}. An alternative method we
found particularly robust to this issue is the Expected Simplex Skewness
(ESS) algorithm proposed by \citet{Johnsson2015}. For example, consider
5000 observations sampled from a \(D=d=50\) dimensional Gaussian
distribution. With the following code, we can see how \texttt{twonn()}
underestimates the true ID, which the ESS instead recovers well.

\begin{verbatim}
R> set.seed(12211221)
R> X_highdim <- replicate(50, rnorm(5000))
R> intrinsicDimension::essLocalDimEst(X_highdim)
\end{verbatim}
\begin{verbatim}
Dimension estimate: 49.05083 
Additional data: ess 
\end{verbatim}
\begin{verbatim}
R> summary(twonn(X_highdim))
\end{verbatim}
\begin{verbatim}
Model: TWO-NN
Method: MLE
Sample size: 5000, Obs. used: 4950. Trimming proportion: 1%
ID estimates (confidence level: 0.95)

| Lower Bound| Estimate| Upper Bound|
|-----------:|--------:|-----------:|
|    34.92154| 35.90791|    36.92254|
\end{verbatim}

However, the ESS is not uniformly optimal. For example, for the
Swissroll data, the \texttt{twonn()} performs better:

\begin{verbatim}
R> intrinsicDimension::essLocalDimEst(Swissroll)
\end{verbatim}
\begin{verbatim}
Dimension estimate: 2.898866 
Additional data: ess 
\end{verbatim}
\begin{verbatim}
R> summary(twonn(Swissroll, c_trimmed = .001))
\end{verbatim}
\begin{verbatim}
Model: TWO-NN
Method: MLE
Sample size: 1000, Obs. used: 999. Trimming proportion: 0.1%
ID estimates (confidence level: 0.95)

| Lower Bound| Estimate| Upper Bound|
|-----------:|--------:|-----------:|
|    1.945607|  2.07005|    2.202571|
\end{verbatim}

When dealing with a dataset characterized by many columns, we suggest
checking the discrepancy between our methods and different competitors.
A marked difference in the results should flag the likelihood-based
findings as less reliable. At this point, a legitimate doubt that may
arise regards the validity of the findings we obtained studying the Alon
dataset. Because of its high number of columns (\texttt{D = 2000}), the
ID we recovered may have been strongly underestimated. To validate our
results, we run the ESS estimator on the Alon dataset, obtaining

\begin{verbatim}
R> intrinsicDimension::essLocalDimEst(data = Xalon)
\end{verbatim}
\begin{verbatim}
Dimension estimate: 7.752803 
Additional data: ess 
\end{verbatim}

which is very close to the estimates obtained with our methods,
reassuring us about our conclusions.

\subsection*{C - Gibbs sampler for \texttt{Hidalgo()}}

The steps of the Gibbs sampler are the following:

\begin{enumerate}
	
	\item Sample the mixture weights according to $$\bm{\pi}|\cdots \sim Dirichlet\left(\alpha_1+
	\sum_{i=1}^{n}\mathds{1}_{z_i=1} ,\ldots,\alpha_K+\sum_{i=1}^{n}\mathds{1}_{z_i=K} \right)$$
	\item Let $\bm{z}_{-i}$ denote the vector $\bm{z}$ without its $i$-th element. Sample the cluster indicators $z_i$ according to:
	\begin{equation*}
\begin{aligned}
	\mathbb{P}\left(z_i=k| \bm{z}_{-i},\cdots \right)\propto \pi_{z_i}
	f\left(\mu_{i},\mathcal{N}_i^{(q)}|z_1,\ldots,z_{i-1},k,z_{i+1},\ldots,z_n, \bm{d}\right)
\end{aligned}
	\end{equation*} 
	We emphasize that, given the new likelihood we are considering, the cluster labels are no longer independent given all the other parameters. Let us define $$\bm{z}_i^k=\left(z_1,\ldots,z_{i-1},k,z_{i-1},\ldots,z_{n}\right).$$ 
	Then, let $N_{z_i}(\bm{z}_{-i})$ be the number of elements in the $(n-1)$-dimensional vector $\bm{z}_{-i}$ that are assigned to the same manifold (mixture component) as $z_i$. 
	Moreover, let $m_{i}^{i n}=\sum_{l} \mathcal{N}_{li}^{(q)} \mathds{1}_{z_{l}=z_{i}}$ be the number of points sampled from the same manifold of the $i$-th observation that have ${x}_i$ as neighbor, and let $n_{i}^{i n}(\bm{z})=\sum_{l} \mathcal{N}_{i l}^{(q)} \mathds{1}_{z_{l}=z_{i}} \leq q$ be the number of neighbors of ${x}_i$ sampled from the same manifold. Then, we can simplify the previous formula, obtaining the following full conditional:
	\begin{equation}
\begin{aligned}
	\mathbb{P}\left(z_i=k| \bm{z}_{-i},\cdots \right)\propto & \frac{\pi_{k} d_{k} \mu_i^{-(d_{k}+1)} }{\mathcal{Z}\left(\zeta, N_{z_{i}=k}(\bm{z}_{-i})+1\right)}\times \left(\frac{\zeta}{1-\zeta}\right)^{n_{i}^{i n}(\bm{z}_i^k)+m_{i}^{i n}(\bm{z}_i^k)}\\
	\times&\left(\frac{\mathcal{Z}\left(\zeta, N_{z_{i}=k}(\bm{z}_{-i})\right)}{\mathcal{Z}\left(\zeta, N_{z_{i}=k}(\bm{z}_{-i})+1\right)}\right)^{N_{z_{i}=k}(\bm{z}_{-i})}.
\end{aligned}
	\end{equation} 
	See \citet{tesiFacco} for a detailed derivation of this result.

	\item The posterior distribution for $\bm{d}$ depends on the prior specification we adopt:
	\begin{enumerate}
\item If we assume a conjugate Gamma prior, we obtain $$d_k|\cdots \sim Gamma\left(a_0+n_k, b_0+\sum_{i:z_i=k}\log \mu_i\right),$$ where $n_k=\sum_{i=1}^{n}\mathds{1}_{z_i=k}$ is the number of observations assigned to the $k$-th group;\\ 
\item If $G_0$ is assumed to be a truncated Gamma distribution on $\left(0,D\right)$, then $$d_k|\cdots \sim Gamma\left(a_0+n_k, b_0+\sum_{i:z_i=k}\log \mu_i\right)\mathds{1}(\cdot)_{\left(0,D\right)};$$
\item   Finally, let us define $a^*=a_0+n_k$ and $b^*=b_0+\sum_{i:z_i=k}\log \mu_i$, if $G_0$ is assumed to be a truncated Gamma with point mass at $D$ we obtain 
$$d_k|\cdots \sim \frac{\hat{\rho}_1^*}{\hat{\rho}_1^*+\hat{\rho}_0^*}\: Gamma\left(a^*, b^*\right)\mathds{1}(\cdot)_{\left(0,D\right)} + \frac{\hat{\rho}_0^*}{\hat{\rho}_1^*+\hat{\rho}_0^*}\:\delta_D(\cdot),$$
where $\hat{\rho}_1^*=\hat\rho\cdot (\mathcal{C}_{a^*,b^*,D}/\mathcal{C}_{a,b,D})$ and $\hat{\rho}_0^*=(1-\hat\rho)\cdot D^{n_k}\cdot\exp\{-D\sum_{i:z_i=k}\log \mu_i\}$.
	\end{enumerate}
\end{enumerate}

\subsection*{D - Postprocessing to address label-switching}

The postprocessing procedure adopted for the raw chains fitted by
\texttt{Hidalgo()} works as follows.

Recall that we are working with \(n\) observations and \(K\) mixture
components. Let us consider an MCMC sample of length \(T\), and denote a
particular MCMC iteration with \(t\), \(t=1,\ldots,T\). Let \(z_i{(t)}\)
indicate the cluster membership of observation \(i\) at the \(t\)-th
iteration, with \(i=1,\ldots,n\). Similarly, \(d_k{(t)}\) represents the
value of the estimated ID in the \(k\)-th mixture component at the
\(t\)-th iteration, where \(k=1,\ldots,K\).\\
We map the \(K\) chains of the parameters in \(\bm{d}\) to each data
point via the values of \(\bm{z}\). That is, we construct \(n\) chains,
one for each observation. At the \(t\)-th iteration, we will have
\(\{d_{z_i{(t)}}{(t)}\}_{i=1}^n\). In so doing, we obtain a collection
of \(n\) chains that link every observation to its ID estimate. When the
chains have been postprocessed, the local observation-specific ID can be
estimated by the ergodic mean or median.

\subsection*{E - The variation of information metric}

This section provides additional details regarding the variation of
information (VI) distance between partitions, a quantity often employed
to estimate optimal posterior clustering configurations.

First, let \(|A|\) denote the cardinality of a generic set \(A\). Then,
consider two different partitions of \(n\) elements defined as
\(\rho_1 = \{S^1_1\ldots,S^1_{p_1}\}\) and
\({\rho_2} = \{S^2_1\ldots,S^2_{p_2}\}\). By definition, for \(q=1,2\),
we have that \(S^{q}_i\cap S^q_j = \varnothing\) when \(i\neq j\), that
\(|\rho_1|=p_1\), \(|\rho_2|=p_2\), and that
\(\sum_{l=1}^{p_q}|S^q_l|=n\). Following the notation in
\citet{Dahl2022}, we define the individual entropy function as
\(H(\rho_q)=-\sum_{S \in \rho_q} |S|/n \log _2\left(|S|/n\right)\) for
\(q=1,2\). Moreover, the joint entropy \(H(\rho_1, \rho_2)\) and mutual
information \(I(\rho_1,\rho_2)\) are defined as

\begin{equation}
	\begin{aligned}
H(\rho_1, {\rho_2}) &=-\sum_{S^1 \in \rho_1} \sum_{S^2 \in {\rho_2}} \frac{|S^1 \cap S^2|}{n} \log _2\left(\frac{|S^1 \cap S^2|}{n}\right), \nonumber\\
I(\rho_1, {\rho_2}) &=H(\rho_1)+H({\rho_2})-H(\rho_1, {\rho_2}).
\nonumber
	\end{aligned}
\end{equation}

Given these quantities, \citet{MeilaM2007, Vinh2010} considered the
variation of information as a distance between partitions:

\begin{equation}
	\mathcal{L}_{VI}(\rho_1, \rho_2) = H(\rho_1) + H(\rho_2) - 2I(\rho_1, \rho_2)
	= -H(\rho_1) - H(\rho_2) + 2H(\rho_1, \rho_2).
	\label{eq:lvi}  
\end{equation}

\citet{Wade2018} used Equation\nobreakspace{}\ref{eq:lvi} as a loss
function to measure the discrepancy between the targeted posterior
partition \(\rho_1=\rho\) and an estimated one \(\rho_2 = \hat{\rho}\).
By minimizing the posterior expectation of
\(\mathcal{L}_{VI}(\rho, \hat{\rho})\) w.r.t. \(\hat{\rho}\), one
obtains the optimal clustering solution:
\[ \hat{\rho}^* = \arg\min_{\hat{\rho}} \mathbb{E}\left[\mathcal{L}_{VI}(\rho, \hat{\rho})\mid \mathcal{D}\right],\]
where \(\mathcal{D}\) denotes the data. Unfortunately, minimizing the
previous quantity (or one of its variations) over the partition space is
extremely challenging. Therefore, numerous authors focused on developing
efficient algorithms for this task: see, for example,
\citet{Wade2018,Rastelli2018} and the review in \citet{Dahl2022}.

\subsection*{F - Global and local intrinsic dimensions}

In Section \ref{sec:hidalgo}, we introduced HIDALGO as a heterogeneous
ID estimator. The Bayesian mixture segments the data into groups, each
belonging to a specific manifold with a specific ID. However, to be more
precise, we provide additional comments on the meaning of the word
\emph{heterogeneous}. In the ID literature, there is a clear distinction
between \emph{global} and \emph{local} ID estimators.\\
On the one hand, methods in the former group estimate a single ID value
for the whole dataset. A single measure for the complexity data is
extremely useful, for example, as a starting point for dimensionality
reduction techniques.\\
On the other hand, the latter group contains methods that attribute a
specific ID to each data point. Such output is instrumental in
monitoring the behavior of the ID across the entire dataset to detect
significant differences in its topology. Moreover, local ID estimation
has an impact when employed for subspace outlier detection, subspace
clustering, or, more in general, other applications in which the ID can
vary from location to location \citep{Amsaleg2019}. Finally, we can
recover a global ID value by aggregating local ID estimates.

Broadly speaking, HIDALGO can be seen as both a global and local
estimator. Its mixture formulation allows the segmentation of the
observations in homogeneous and spatially related model-based clusters.
Hence, we can see HIDALGO as an estimator for multiple global IDs (one
for each manifold but not for each point). Nonetheless, we can take
advantage of the Bayesian framework and the postprocessing procedure we
propose to deal with the label-switching issue (see Section D of the
Appendix for more details). Indeed, while solving for label-switching,
our procedure delivers a valuable byproduct. By mapping the \(K\)
mixture parameters \(\{ d_k \}_{k=1}^K\) into \(n\) different
observation-specific chains, we can effortlessly obtain an ID value for
each data point. For example, see Figure \ref{fig:memed}, where
observation-specific posterior mean and median ID estimates are
displayed.

\subsection*{G - System configuration}

We obtained the results in this vignette by running our \texttt{R}
code on a MacBook Pro with a 2.6 GHz 6-Core Intel Core i7 processor.

\bibliographystyle{plainnat}
\bibliography{main}

\end{document}